 \documentclass[final,5p,times,twocolumn]{elsarticle}

  \usepackage{graphicx}
  \usepackage{epstopdf}
  \usepackage{verbatim}
  \usepackage{subfigure}

\usepackage{amssymb}

\begin{document}

\begin{frontmatter}



\title{Shallow water rogue wavetrains in nonlinear optical fibers}


 \author[Brescia]{Stefan Wabnitz}
 \ead{stefano.wabnitz@ing.unibs.it}
 \author[Dijon]{Christophe Finot}
 \author[Dijon]{Julien Fatome}
 \author[Dijon]{Guy Millot}

 \address[Brescia]{Dipartimento di Ingegneria dell'Informazione,  Universit\`a degli Studi di Brescia, via Branze 38, 25123, Brescia, Italy}
 \address[Dijon]{Laboratoire Interdisciplinaire Carnot de Bourgogne, UMR 6303 CNRS, Universit\'e de Bourgogne, 9 av. Alain Savary, 21078 Dijon, France}

\begin{abstract}
In addition to deep-water rogue waves which develop from the modulation instability of an optical CW, wave propagation in optical fibers may also produce shallow water rogue waves. These extreme wave events are generated in the modulationally stable normal dispersion regime. A suitable phase or frequency modulation of a CW laser leads to chirp-free and flat-top pulses or flaticons which exhibit a stable self-similar evolution. Upon collision, flaticons at different carrier frequencies, which may also occur in wavelength division multiplexed transmission systems, merge into a single, high-intensity, temporally and spatially localized rogue pulse. 
\end{abstract}


\end{frontmatter}


\section{Introduction}
\label{sec:intro}
The dynamics of extreme waves, often known as freak or rogue
waves, is presently a subject of intensive research in several fields of 
application \cite{dhyste2008}-\cite{akhmediev2010sp}. In oceanography, rogue waves are mostly known as a sudden deep-water event which is responsible for
ship wreakages. A relatively less explored, but potentially even more damaging manifestation of rogue waves also occurs in shallow waters, consider for example the propagation of tsunamis. In such environment, the crossing of waters propagating in different directions may lead to the formation of high-elevation and steep humps of water that result in severe coastal damages \cite{soomere}. A universal model for describing the formation of deep-water rogue waves is provided by the one-dimensional
Nonlinear Schr\"odinger Equation (NLSE). In this framework, rogue waves are linked with the presence of modulation instability (MI) \cite{benjamin}, whose nonlinear development is described by the so-called Akhmediev breathers \cite{akhmediev86}, and may ultimately result in the formation of the Peregrine soliton, a wave of finite extension in both the evolution and the transverse coordinates \cite{peregrine83}.

An ideal testbed for the experimental study of rogue waves is provided by optical pulse propagation in nonlinear optical fibers, which is closely described by the NLSE. Indeed, the statistics of spectral broadening in optical supercontinuum generation has been associated with extreme solitary wave emissions \cite{solli2007}. Moreover, the first experimental observation of the Peregrine solitons in any physical medium has been carried out exploiting the induced MI occuring in a highly nonlinear fiber \cite{kibler2010}. 

In this Letter, we show that rogue waves in optical fibers may also be generated in the normal group-velocity dispersion (GVD) regime of pulse propagation, where MI is absent. Indeed, nonlinearity driven pulse shaping in this case may be described in terms of the semiclassical approximation to the NLSE \cite{yuji95}, which leads to the so-called nonlinear shallow water equation (NSWE) \cite{whitham}, which is also known in hydraulics as the Saint-Venant equation \cite{stvenant}. Therefore we establish a direct link between the dynamics of extreme wave generation in shallow waters \cite{didenkulova} and their direct counterparts in optical communication systems. Since the CW state of the field is stable, shallow water optical rogue waves may only be generated as a result of particular setting of the initial or boundary conditions. Namely, as discussed by Kodama and Biondini \cite{yuji99}-\cite{biondini06}, the initial modulation of the optical frequency, which is analogous to considering the collision between oppositely directed currents near the beach, or the merging of different avalanches falling from a mountain valley. 

In section \ref{sec:theo} we shall describe the dynamics of the generation of an intense, flat-top, self-similar and chirp-free pulse as a result of the initial step-wise frequency modulation of a CW laser. In hydrodynamics, this corresponds to the hump of water which is generated by two water waves traveling with opposite velocities. The intriguing property of such pulses, that we name flaticons, is their stable merging upon mutual collision into either a steady or transient high-intensity wave, as discussed in section \ref{sec:colli}. The pulse collision dynamics may also lead to the formation of extreme intensity peaks in optical communication systems whenever various wavelength channels are transported on the same fiber. As pointed out in section \ref{sec:wave}, an interesting application of optical shallow water rogue waves is the possibility of generating, from a frequency or phase modulated CW laser, high repetition rate pulse trains with low duty ratio. In contrast with existing linear techniques for pulse train generation \cite{koba88}-\cite{otsuji96}, rogue wavetrains lead to chirp-free, high intensity pulse trains, which are important advantages for their possible use as communication signals.       

\section{Optical pulse dynamics}
\label{sec:theo}

The propagation of pulses in optical fibers is described by the NLSE

\begin{equation}
	\nonumber i\frac{\partial Q}{\partial z}-\frac{\beta_2}{2}\frac{\partial^2 Q}{\partial t^2}+\gamma|Q|^2Q=0.\label{nls1}
\end{equation}

\noindent Here $z$ and $t$ denote the distance and retarded time (in the frame travelling at the group-velocity) coordinates;  $\beta_2$ and $\gamma$ are the group velocity dispersion and the nonlinear coefficient, and $Q$ is the field envelope. In dimensionless units, and in the normal dispersion regime (i.e., $\beta_2>0$), Eq.(\ref{nls1}) reads as

\begin{equation}
	\nonumber i\frac{\partial q}{\partial Z}-\frac{\beta^2}{2}\frac{\partial q}{\partial T^2}+|q|^2q=0.\label{nls2}
\end{equation}

\noindent where $T=t/t_0$, $Z=z\gamma P_0=z/L_{NL}$, $\beta^2=\beta_2/(T_0^2\gamma P_0)\equiv L_{NL}/L_D$, where $L_{NL}$ and $L_D$ are the nonlinear and dispersion lengths, respectively, $q=Q/\sqrt{P_0}$ , $t_0$ and $P_0$ are arbitrary time and power units. Eq.(\ref{nls2}) can be expressed in terms of the real variables $\rho$  and $u$ which denote the field dimensionless power and instantaneous frequency (or chirp)

\begin{equation}
	\nonumber q(T,Z)=\sqrt{\rho(T,Z)}\exp\left[-\frac{i}{\beta}\int_{-\infty}^{T}u(T',Z)dT'\right].\label{ansatz}
\end{equation}
 
By ignoring higher order time derivatives in the resulting equations (which is justified for small values of $\beta$), one obtains from the NLSE the semi-classical or hydrodynamic NSWE \cite{yuji95}-\cite{whitham}

\begin{equation}\label{swe}
\frac{\partial }{\partial Z'} \left ( \begin{array} {l}
\rho \\ u \end{array}\right) + \left ( \begin{array} {cc}
u & \rho \\ 1 & u \end{array}\right)\frac{\partial }{\partial T}\left ( \begin{array} {l}
\rho \\ u \end{array}\right)=0,
\end{equation}

\noindent where $Z'=\beta Z$. In hydrodynamics, Eq.(\ref{swe}) describes the motion of a surface wave in shallow water, i.e., a wave whose wavelength is much larger than the water depth. In this context, 
$\rho$ and $u$ represent the water depth and its velocity, respectively. For a temporally localized input optical waveform such as a chirp-free square pulse (which is representative of the nonreturn-to-zero (NRZ) optical modulation format), i.e., with $\rho(T,Z=0)=\rho_0$ for $\left|T\right|\leq T_0$ and $\rho(T,Z=0)=0$ otherwise, Eq.(\ref{swe}) may be analytically solved up to the point $Z'=T_0/\sqrt{\rho_0}$ in terms of the well-known Ritter dam-break solution \cite{yuji95},\cite{ritter}. Note that, at this point, the initial square NRZ pulse has broadened into a triangular pulse. In order to counteract such pulse deformation, it was proposed in \cite{yuji95b} to use an input step-wise periodic frequency modulation, so that the self-phase modulation-induced chirp can be largely compensated for.

\begin{figure}[ht]
\centering
\includegraphics[width=6cm]{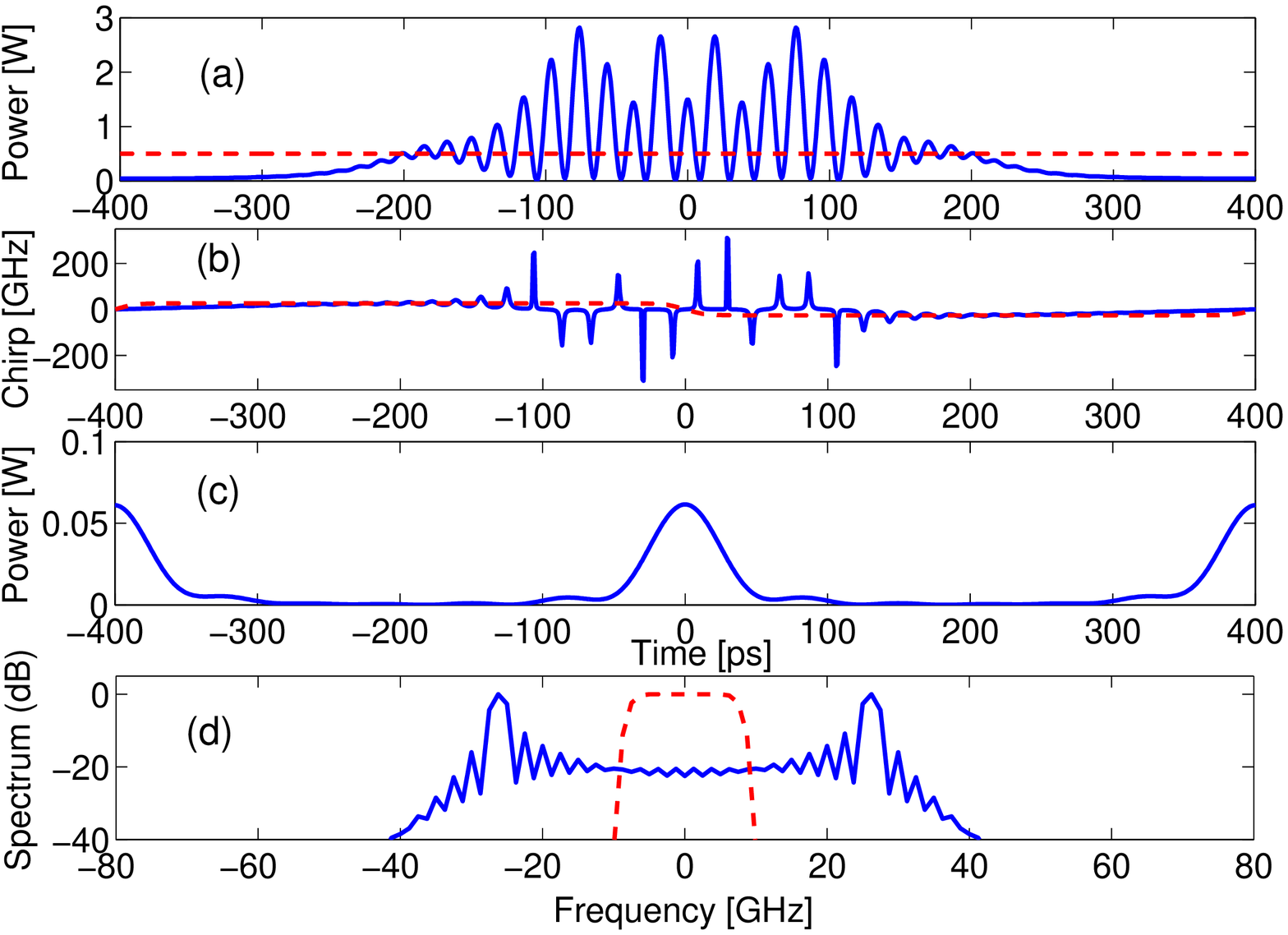}
\includegraphics[width=6cm]{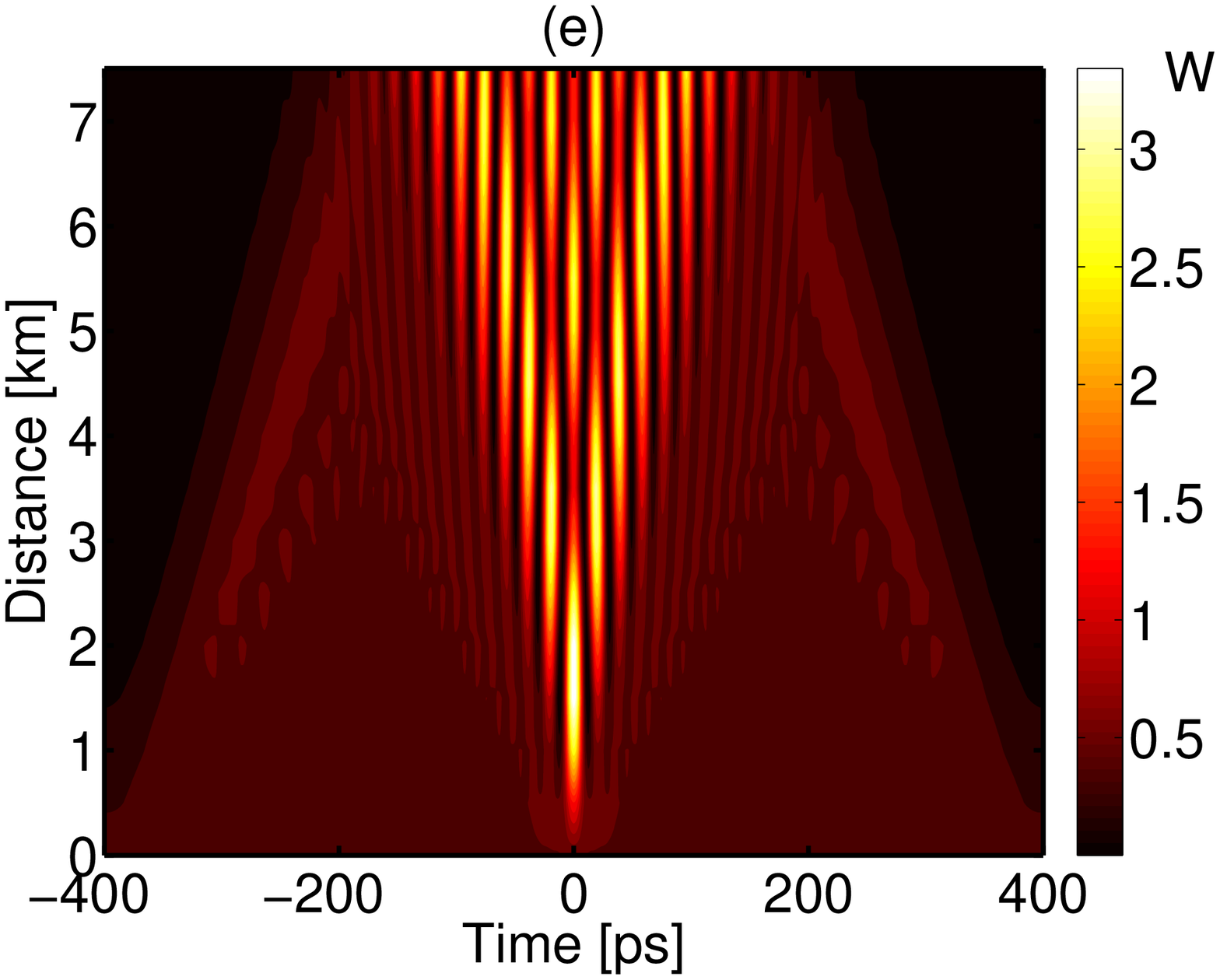} 
\caption{Output (blue solid curves) and input (red dashed curves) power (a) and chirp (b) profiles from a 7 km long DCF in the linear case (i.e., with $\gamma=0$); (c) output pulse power after a bandpass filter with 15.5 GHz bandwidth; (d) input and output spectral intensity (solid blue curve) and intensity transmission function of bandpass filter (dashed red curve); (e) contour plot of power profile vs. length.}
\label{sjumplin}
\end{figure}

We are interested here in studying the behavior of the solutions of Eq.(\ref{swe}) with a dual quasi-CW pump input, that is we set $\rho(T,Z=0)=\rho_0,\forall T$ , and a periodic (with period $T_M$) frequency modulation, namely 

\begin{equation}\label{fjump}
	u(T,Z=0)=\left\{\begin{array} {l}
u_0 \: for \:-T_M/2<T<0\\ -u_0 \: for \:0<T<T_M/2 \end{array}\right. ,
\end{equation}

\noindent so that in each modulation period there are two opposite frequency jumps. Indeed Eq.(\ref{fjump}) corresponds to the injection of two time alternating, quasi-CW pumps at opposite frequencies $\pm u_0$, respectively (see the power spectrum in panel (d) of Fig.\ref{sjumplin}). Supposing that $u_0>0$, the frequency jump at $T=0$ is such that, because of normal dispersion, the leading wave components at $T<0$ travel slower than the trailing components at $T>0$. Hence a wave compression (optical piston effect) at $T=0$ results, which leads to a dispersive shock or optical wave breaking. That is, high-frequency oscillations appear with a characteristic oscillation frequency equal to $1/\beta$. The opposite situation occurs for the frequency jump at $T=\pm T_M/2$, where dispersion leads to wave rarefaction, so that a dark pulse or hole develops.

Indeed, as shown in panel (a) of Fig.\ref{sjumplin}, high-intensity oscillations also occur in a purely linear dispersive medium (i.e., whenever $\gamma=0$ in Eq.(\ref{nls1})), owing to the beating among the different frequency components which are generated by the initial condition Eq.(\ref{fjump}), and that travel at different speeds. Here we show the output power profile from a 7 km long dispersion-compensating fiber (DCF) with normal GVD $D=-100 ps/(nm\cdot km)$ (or $\beta_2=127 ps^2/km$) under purely linear propagation conditions. In Fig.\ref{sjumplin} we considered a 1.25 GHz rate of frequency modulation with $\pm26\:GHz$ amplitude, and the input CW power $P=500\: mW$ (or 27 dBm).  

Panel (b) of Fig.\ref{sjumplin} shows that in the linear case the output wavetrain develops a strong chirp as it propagates. Clearly in the absence of nonlinearity the spectrum of panel (d) in Fig.\ref{sjumplin} remains unchanged, with most of its energy concentrated at the two quasi-CW pump frequencies. Therefore if we place a relatively narrow bandpass filter centered at the carrier frequency $u=0$, we only obtain a weak (i.e., with a peak power which remains two orders of magnitude lower than the peak of the oscillations in panel (a) of Fig.\ref{sjumplin}) periodic pulse train (see panel (c) of Fig.\ref{sjumplin}). We used here a filter with the supergaussian spectral amplitude transfer function of order 20, with spectral bandwidth of 15.5 GHz, calculated at -3dB from the peak transmission value. The filter transfer function is illustrated by a red dashed curve in panel (d) of Fig.\ref{sjumplin}, and compared to the constant spectral intensity profile (solid blue curve). Finally, panel (e) of Fig.\ref{sjumplin} shows that the width of the region with temporal oscillations grows larger with distance owing to linear dispersive spreading. 

\begin{figure}[ht]
\centering
\includegraphics[width=6cm]{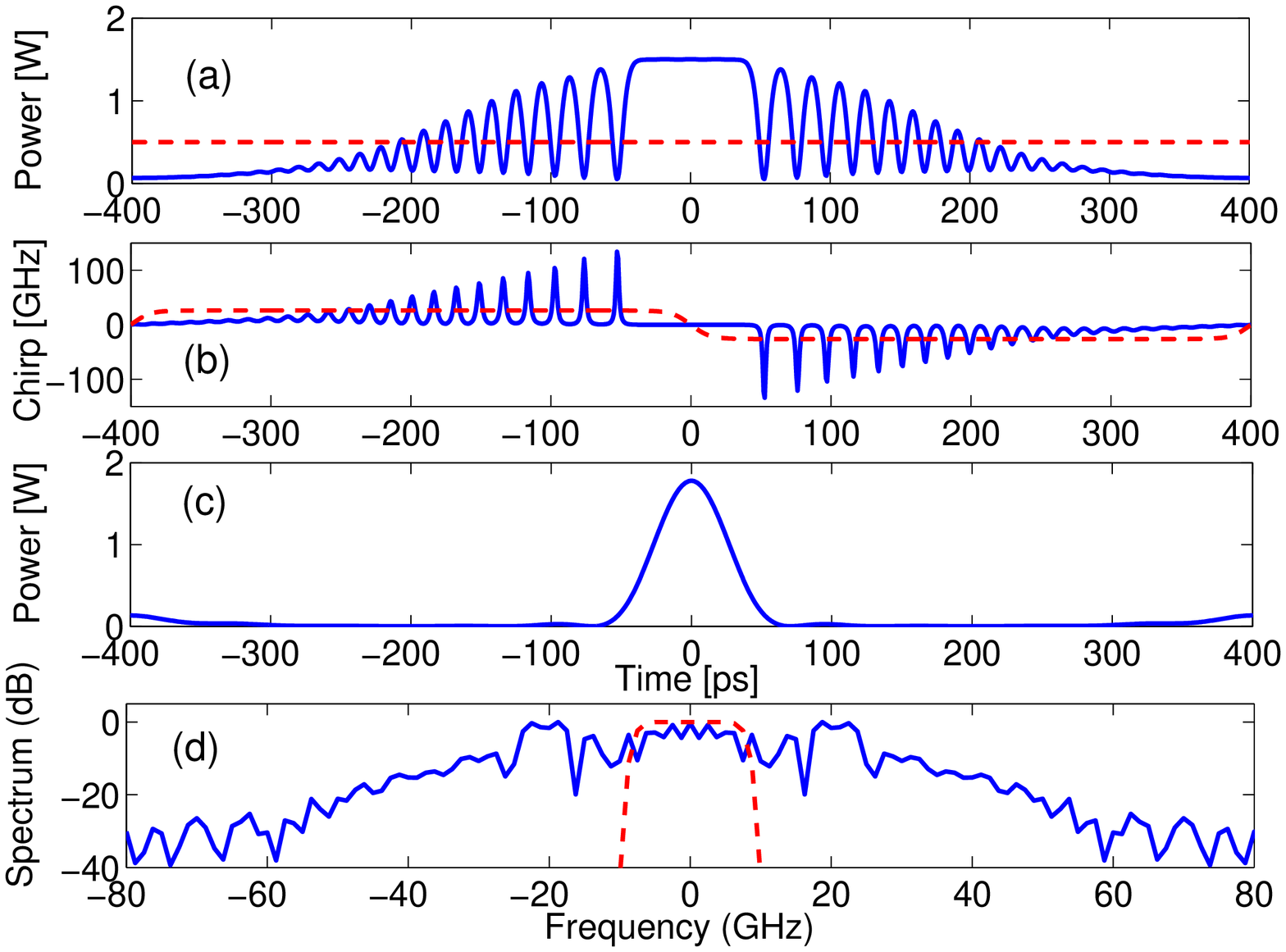}
\includegraphics[width=6cm]{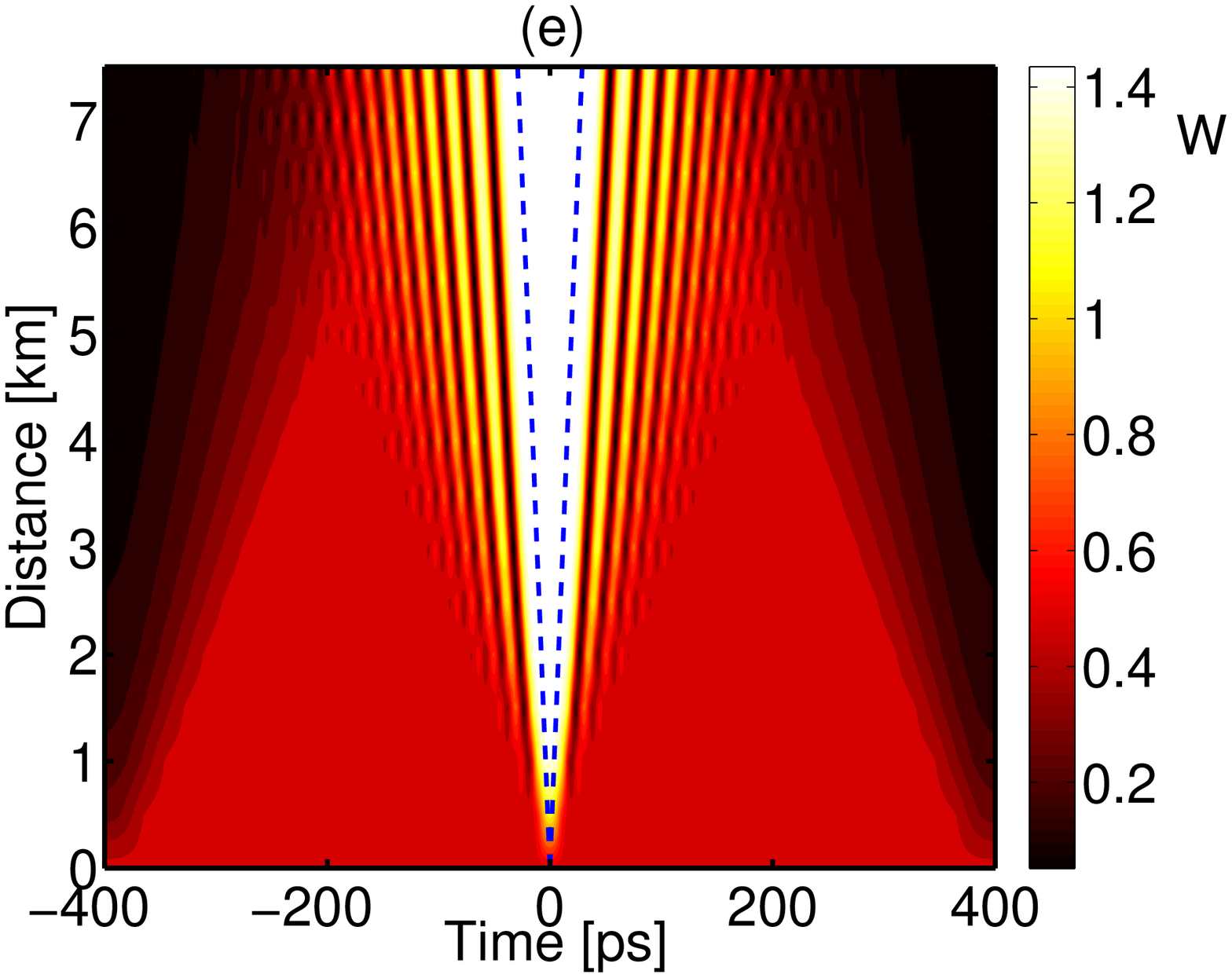} 
\caption{Output (blue solid curves) and input (red dashed curves) power (a) and chirp (b) profiles from a 7 km long DCF; (c) output pulse after a bandpass filter with 15.5 GHz bandwidth; (d) output spectral intensity (solid blue curve) and intensity transmission function of bandpass filter (dashed red curve); (e) contour plot of flaticon intensity profile vs. DCF length, and analytical prediction (\ref{tp}) of flaticon edges (dashed blue curves).}
\label{sjump}
\end{figure}

In the nonlinear case (e.g., with the nonlinear coefficient $\gamma =3.2 W^{-1}km^{-1}$ in Eq.(\ref{nls1})), wave propagation is dramatically different. On the one hand, the collision among the two oppositely traveling (in the reference frame traveling with the group velocity of the carrier component at $u=0$) quasi-CW pumps leads to a dispersive wave-breaking or shock, as it is illustrated in panel (a) of Fig.\ref{sjump}. 

On the other hand, panel (a) of Fig.\ref{sjump} also shows that, in spite of the presence of the temporal shock, whenever the magnitude of the initial chirp $u_0$ remains below a critical value, say, $u_c$, i.e., if $u_0\leq u_c\equiv 2\sqrt{\rho_0}$, there is a finite time interval across the frequency jump at $T=0$ where a high-power and chirp-free pulse is generated \cite{yuji99}-\cite{biondini06}. Such a pulse exhibits a flat-top, nonoscillating behavior, hence we may name it a "flaticon". The flaticon is a stable waveform, as it maintains a self-similar shape with distance: Kodama and Biondini \cite{biondini06} have analytically demonstrated that its temporal duration, say, $T_p$, obeys the simple linear law

\begin{equation}\label{tp}
	T_p=\left(2\sqrt{\rho_0}-u_0\right)Z'\equiv2V_pZ'
\end{equation}
 
\noindent where $\pm V_p$ is the speed of propagation of the flaticon edges. In panel (e) of Fig.\ref{sjump} we compare  the prediction of Eq.(\ref{tp}) with the numerical solution of the NLSE (\ref{nls2}). As it can be seen, Eq.(\ref{tp}) correctly captures the self-similar expansion of the flaticon temporal duration with the propagation distance. The flaticon time duration in Fig.\ref{sjump} is slightly longer than the value of Eq.(\ref{tp}), owing to the finite width of the initial smooth frequency transition as opposed to the sudden jump in Eq.(\ref{fjump}).

In real units, the critical frequency shift $f_c$ reads as

\begin{equation}\label{fcrit}
	f_c=\frac{u_c}{2\pi t_0\beta}=\frac{1}{\pi t_0\beta}\sqrt{\frac{P}{P_0}}=\frac{1}{\pi}\sqrt{\frac{\gamma P}{\beta_2}}
\end{equation}

\noindent Eq.(\ref{fcrit})) shows that the amplitude of the critical frequency modulation is controlled by varying the ratio of the input power $P$ to the fiber GVD $\beta_2$. On the other hand, the validity of the condition $\beta^2<<1$ may be independently ensured by a suitable choice of the reference time $t_0$.
 
Quite strikingly, panel (e) of Fig.\ref{sjump} shows that the flaticon power remains constant with distance. In fact, its peak power is equal to the value

\begin{equation}\label{rop}
	\rho_p=\rho_0\left(1+\frac{u_0}{2\sqrt{\rho_0}}\right)^2
\end{equation}

\noindent which has a maximum value of $\rho_c=4\rho_0$  for $u=u_c$. However in the critical case the flaticon temporal duration $T_p$ shrinks to zero. In Fig.\ref{sjump} we have set $u_0=(\sqrt{3}-1)u_c$, which leads to $\rho_p=3\rho_0$. Eqs.(\ref{tp})-(\ref{rop}) mean that the flaticon energy grows larger linearly with distance, as it continuosly draws energy towards the zero-frequency region, at the expense of the two input quasi-CW waves at frequencies $\pm u_0$.

In order to extract the flaticon from the high-frequency pumps and from its highly chirped oscillating tails, it is necessary to use a suitably optimized bandpass filter which highly increases the extinction ratio of the generated optical pulse, without sacrificing its peak power, as it can be seen by comparing panels (a) and (c) of Fig.\ref{sjump}. Here we used the same bandpass filter as in the case of Fig.\ref{sjumplin}.

\section{Rogue wave collisions}
\label{sec:colli}

As we have seen, the peak power of the chirp-free flaticon pulse which results from a square-wave frequency modulation of amplitude $2u_0$ (see Eq.(\ref{fjump})) has a maximum value of $\rho_{c}=4\rho_0$, which is obtained for the critical value $u_0=u_c=2\sqrt{\rho_0}$. However this limit case is not interesting in practice, since the temporal width of the pulse shrinks to zero. Nevertheless, as pointed out in \cite{biondini06}, it is remarkable that a finite width flaticon pulse with peak power  $\rho_c$ can still be obtained as a result of the collision and merging of two flaticons travelling at different carrier frequencies. Such collision may obtained by imposing at least two or more frequency jumps to an input frequency modulation with a total (or collective) amplitude which remains close to $2u_c$.

\begin{figure}[ht]
\centering
\includegraphics[width=6cm]{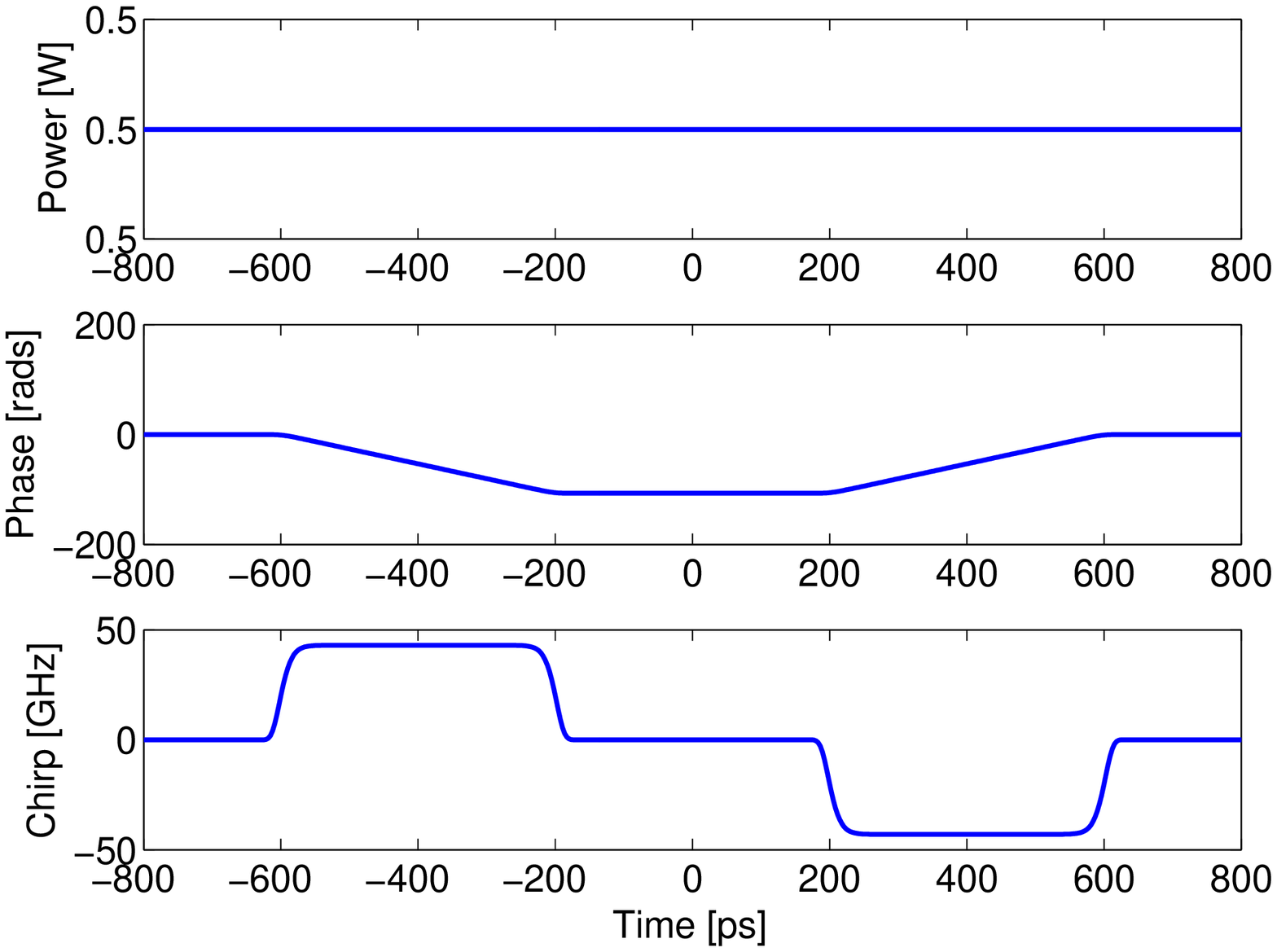}
\includegraphics[width=4cm]{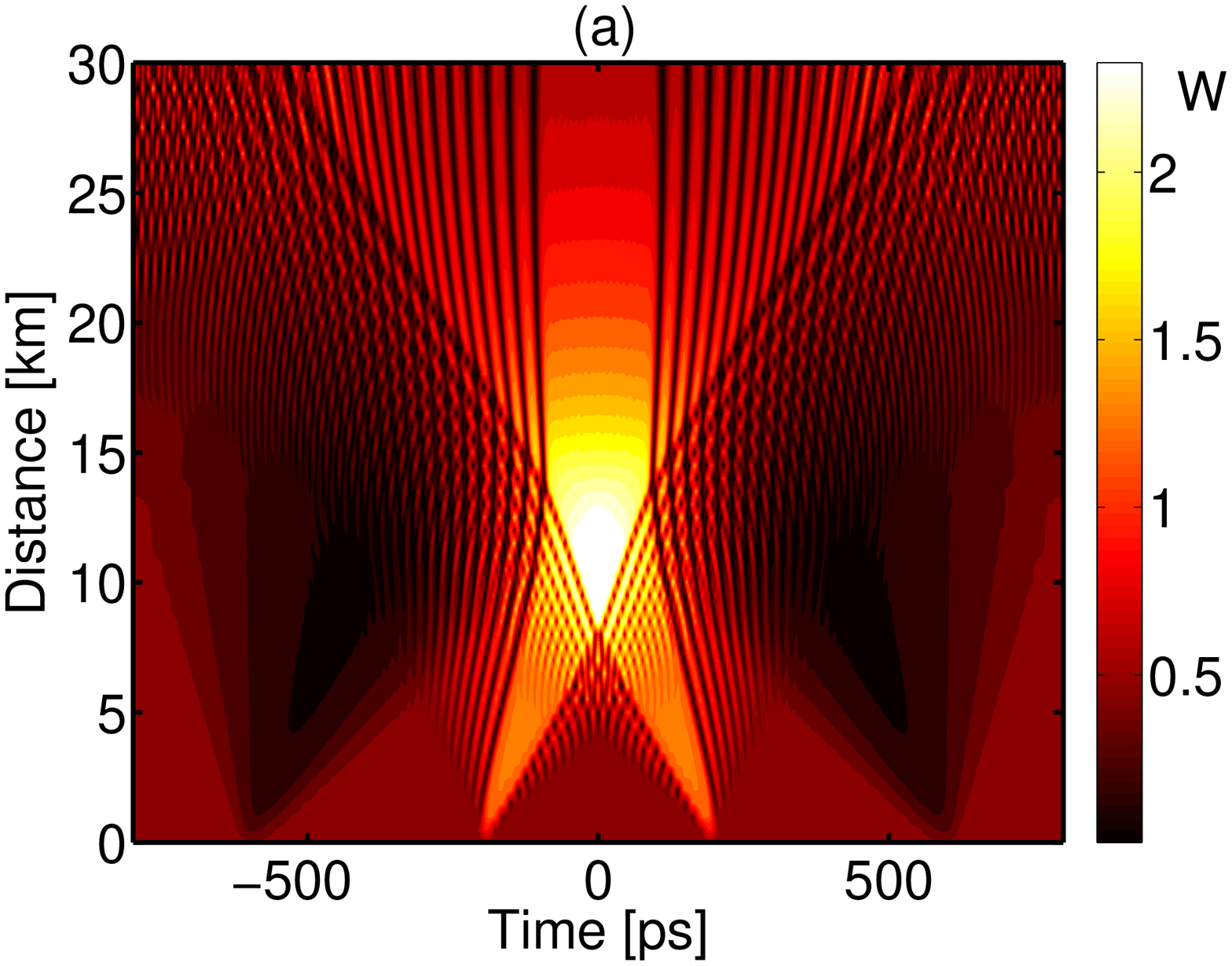}
\includegraphics[width=4cm]{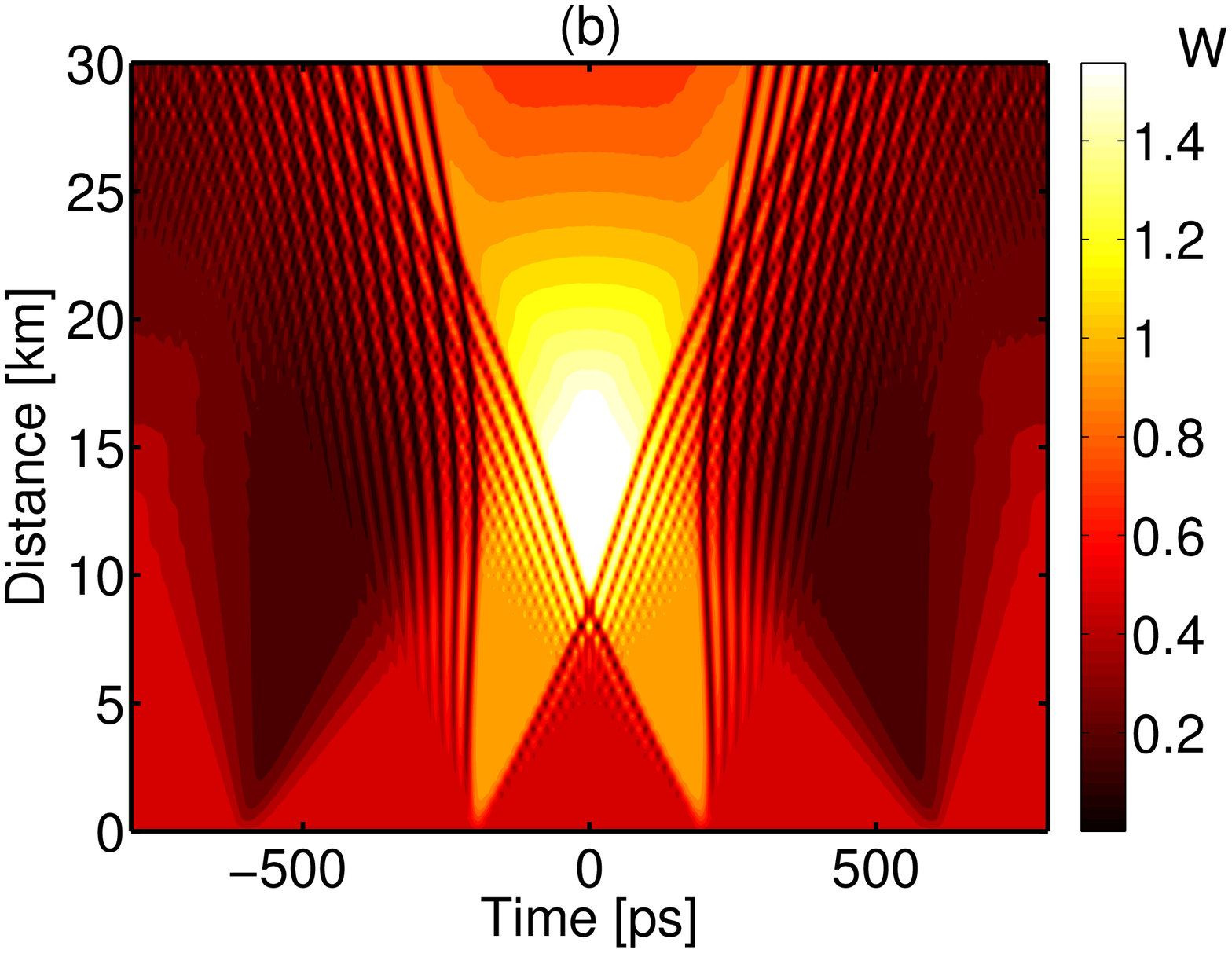} 
\includegraphics[width=4cm]{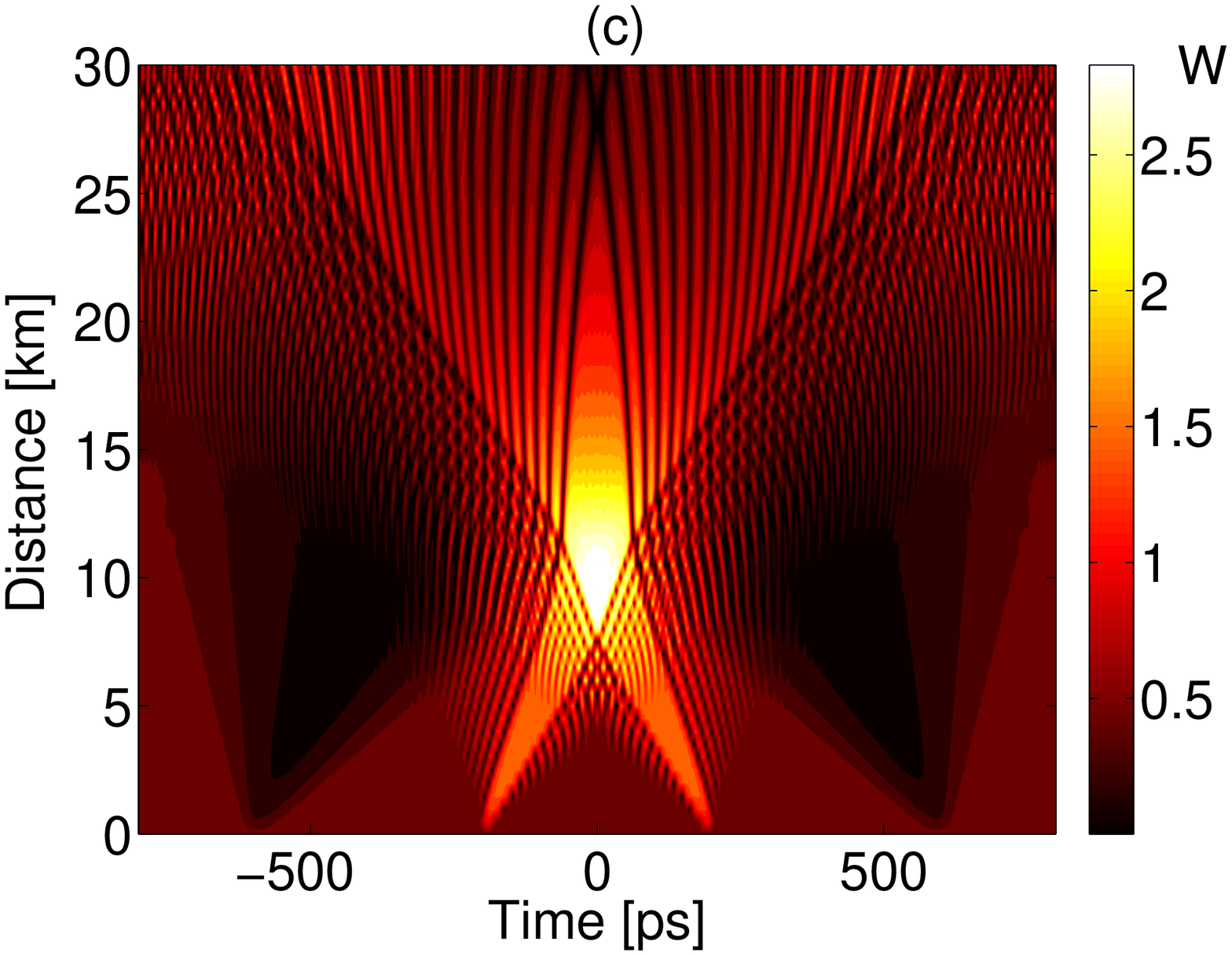}  
\caption{Top: input power, phase and frequency modulation vs. time for the case of two consecutive critical frequency jumps; Panels (a-c): resulting flaticon collisions: (a) critical jump ($f_j=2.4f_c$); (b) subcritical jump ($f_j=1.6f_c$); (c) supercritical jump ($f_j=2.8f_c$).}
\label{coll2}
\end{figure}

Henceforth we will consider frequency modulations which are symmetric around $T=0$, composed by a set of equal amplitude, decreasing frequency jumps. Consider for example the initial frequency modulation (and associated phase modulation) profile that is illustrated in Fig.\ref{coll2}. In this case, each frequency step has the amplitude $f_s=1.2f_c=43\: GHz$, so that a full jump amplitude (between $t=-400\: ps$ and $t=+400\: ps$) of $f_j=2f_s=2.4f_c=86\: GHz$ results. In Fig.\ref{coll2} the time separation among subsequent frequency jumps is equal to 400 ps.

The corresponding contour plot of the total power evolution with distance in the DCF is illustrated in panel (a) of  Fig.\ref{coll2}. As it can be seen, each frequency step has a subcritical amplitude (i.e., $f_s<2f_c$), so that two individual flaticon pulses of the type described by Eqs.(\ref{tp})-(\ref{rop}) are generated. These pulses travel in opposite directions because of the shift (with respect to the central carrier frequency) of their individual mid-frequency. As the two flaticons collide, their nonlinear interaction leads to their merging into a single, higher amplitude flaticon whose temporal edges remain virtually unchanged upon further propagation. Such a situation is described by saying that the total frequency jump has a critical amplitude. 

Note that, by numerically solving the NLSE (\ref{nls2}), we obtain the critical jump amplitude $f_j=2.4f_c$, which is slightly larger than the theoretically predicted value $2f_c$ resulting from the analytical solution of Eq. (\ref{swe}). This is likely due to the fact that we used relatively smooth frequency transitions, and to the relatively large value of $\beta^2=0.2$. 

On the other hand, whenever the total frequency jump amplitude is less than the critical value or subcritical, upon collision the two flaticons merge into a single flaticon with progressively expanding temporal edges (see panel (b) of Fig.\ref{coll2}, where $f_j=1.6f_c$). Finally, for a supercritical (or larger than the critical cumulated value) double frequency jump, the collision leads to a high amplitude and spatially localized sneaker flaticon, whose edges shrink with distance until it disappears after about 30 km (see panel (c) of Fig.\ref{coll2}, where $f_j=2.8f_c$). Therefore the critical frequency jump case in panel (a) of Fig.\ref{coll2} represents a separatrix among the subcritical expanding and the supercritical shrinking cases: the resulting flaticon has a stable steady-state temporal duration although its peak power is slowly decreased with distance.

 Note that, in stark contrast with the collision of NLSE solitons in the anomalous GVD regime, which emerge unchanged from the collision except for a temporal and a time shift, in the normal GVD regime the collisions of subcritical flaticon waves leads to their full merging into a different, high-amplitude sneaker wave which entirely captures their energy.  

Quite remarkably, the peak power of the collision-induced sneaker flaticon remains in good agreement with Eq.(\ref{rop}), where $u_0$ is equal to the semi-amplitude of the total frequency jump. For the critical case in panel (a) of Fig.\ref{coll2} one predicts a peak sneaker flaticon power of $2.4 W$, which is close to the numerically observed value of $2.3 W$ (see the legend of the power values in the (a) panel of Fig.\ref{coll2}). In the subcritical (supercritical) case, Eq.(\ref{rop}) predicts a peak sneaker wave power of $1.6 W$ and $2.9 W$, respectively, again in good agreement with the numerical values observed in the corresponding legends in panels (b) and (c) of Fig.\ref{coll2}.

\begin{figure}[ht]
\centering
\includegraphics[width=6cm]{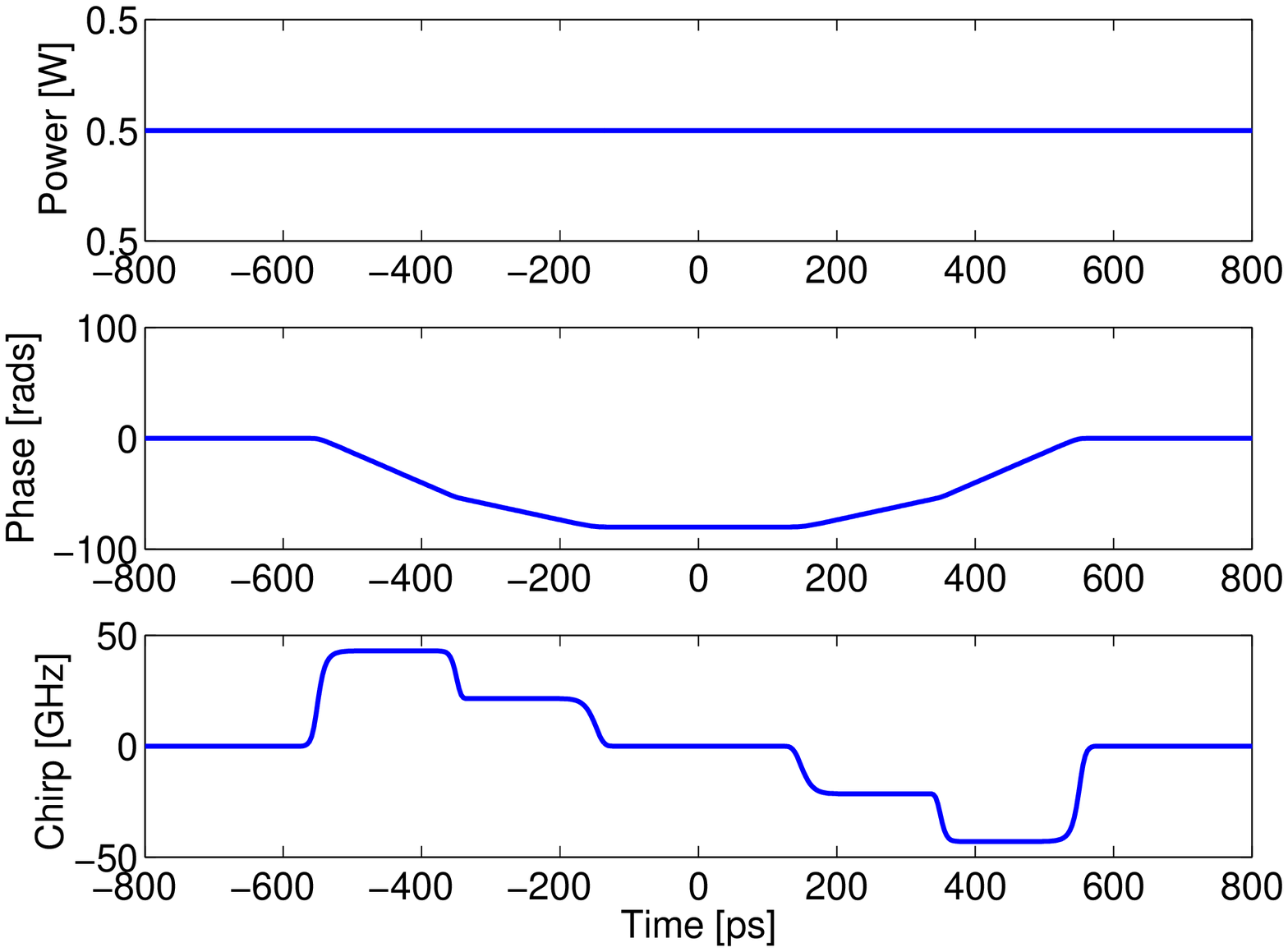}
\includegraphics[width=4cm]{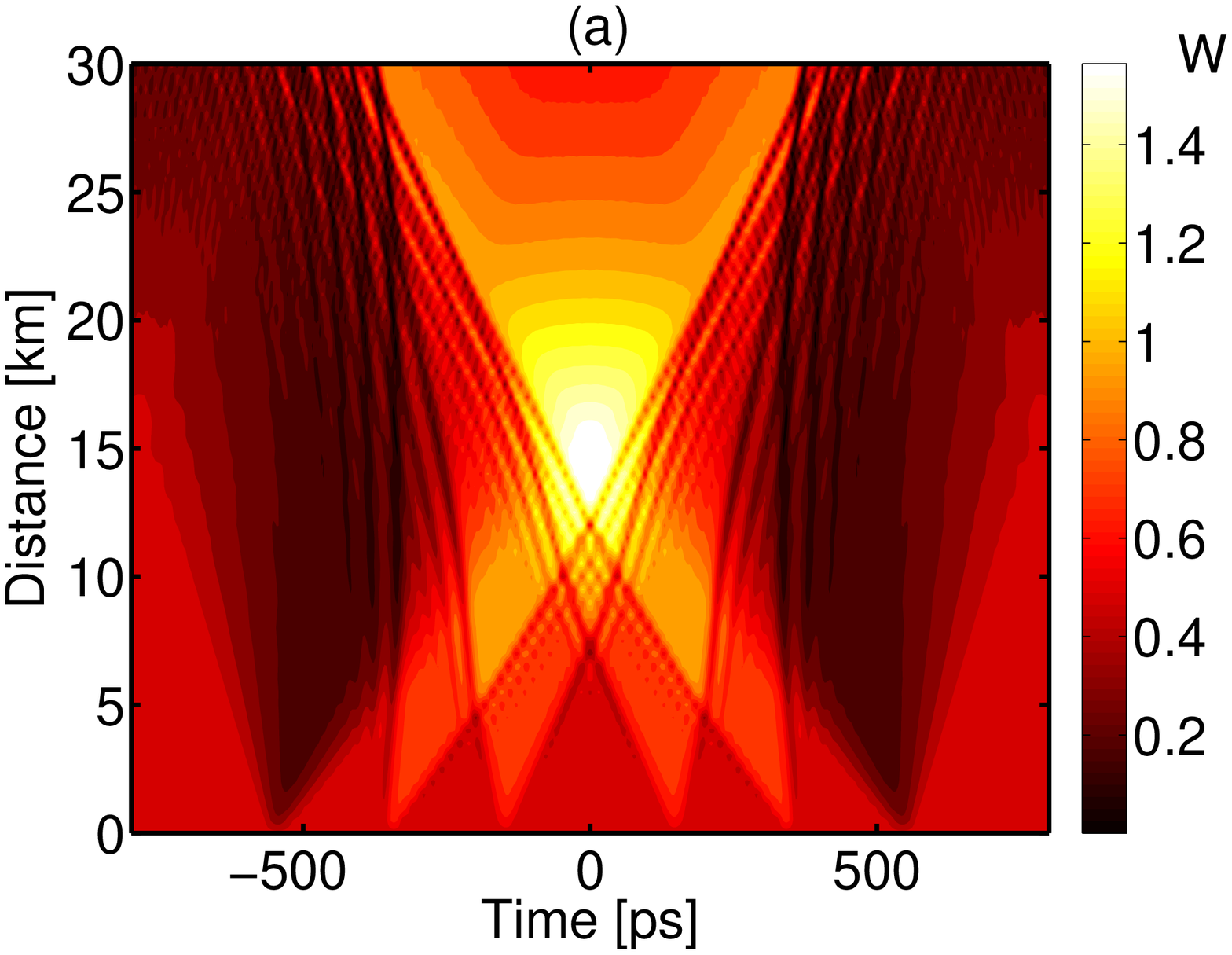}
\includegraphics[width=4cm]{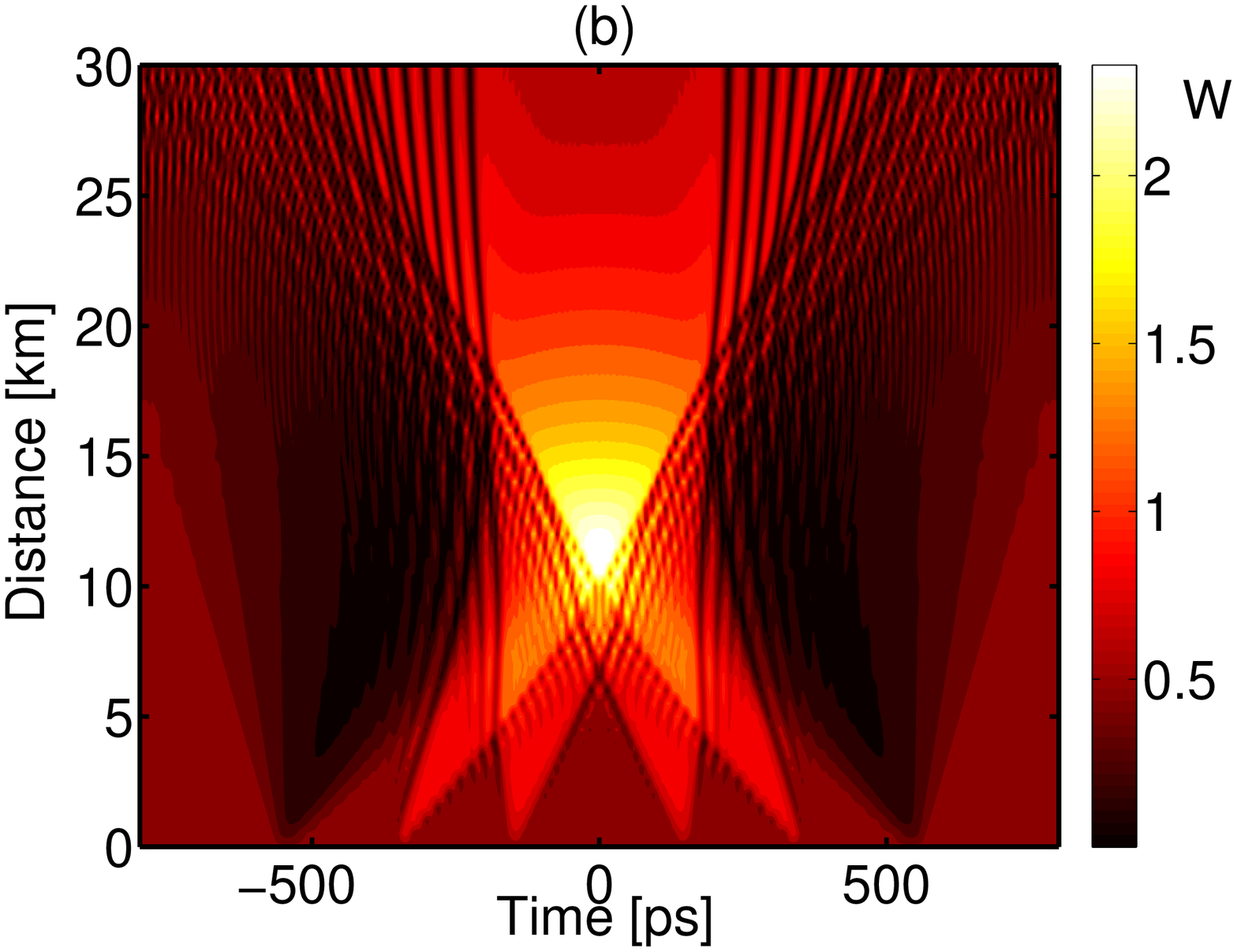} 
\includegraphics[width=4cm]{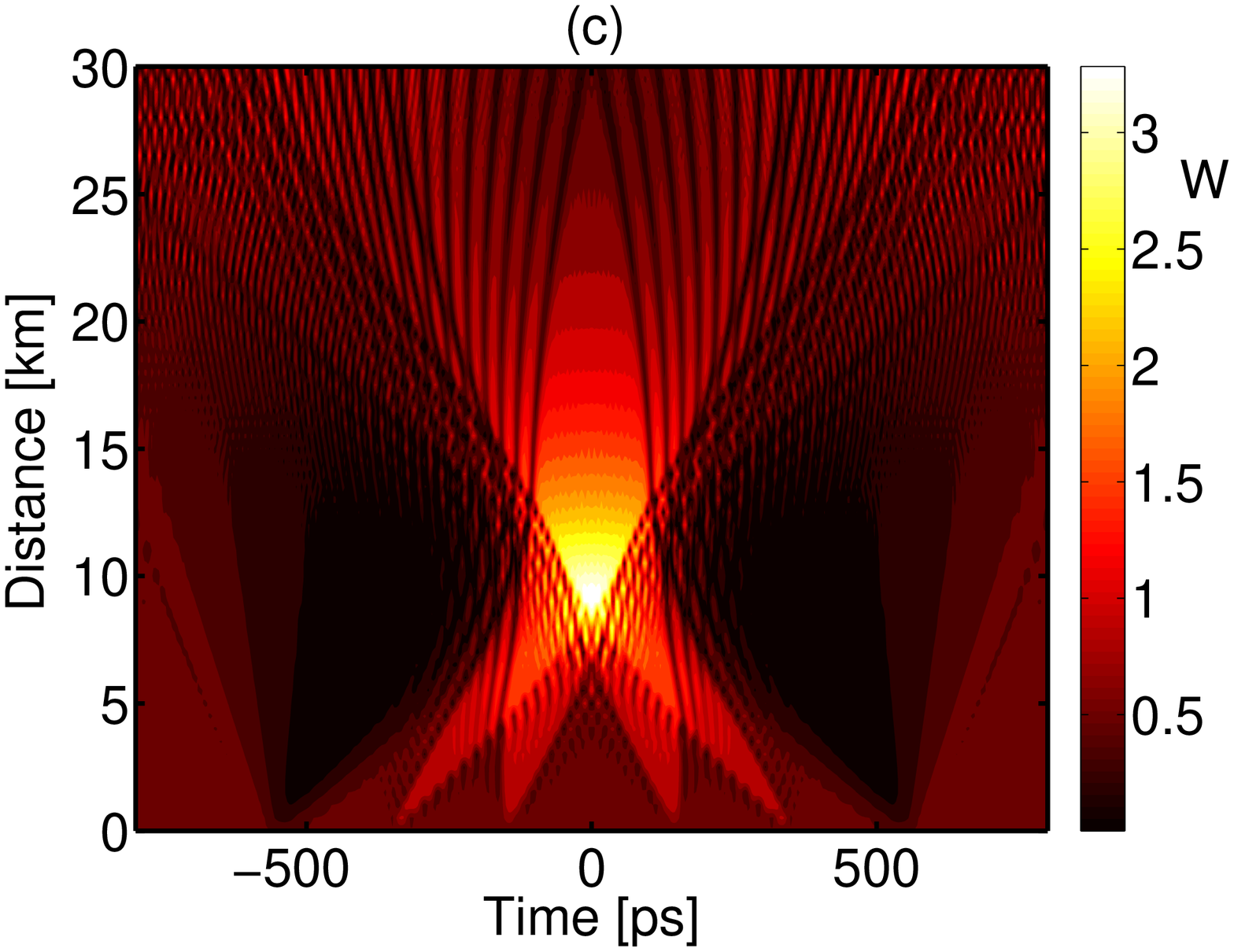}  
\caption{Input power, phase and frequency modulation vs. time for the case of four consecutive critical frequency steps; Panels (a-c): resulting flaticon collisions: (a) Critical total jump ($f_j=2.4f_c$); (b) subcritical total jump ($f_j=1.6f_c$); (c) supercritical total jump ($f_j=3.2f_c$).}
\label{coll4}
\end{figure}

A similar situation occurs in the collision of more than two flaticons: in Fig.\ref{coll4} we illustrate the case of four-pulse collisions, resulting from four consecutive frequency down steps. Since the critical multiple frequency jump always corresponds to a total frequency modulation amplitude close to the critical value $2f_c$, in Fig.\ref{coll4} the amplitude of each step has been divided by two with respect to the case of Fig.\ref{coll2}, namely $f_s=0.6f_c=21.5\: GHz$, so that a collective jump amplitude of $f_j=4f_s=2.4f_c=86\: GHz$ is maintained. In Fig.\ref{coll4}, the frequency jumps occur at $t=\pm 350 ps$  and $t=\pm 150 ps$, respectively, so that the collision among the four flaticons occurs at approximately the same distance as in Fig.\ref{coll2}. Note that the initial time separation among the colliding flaticons does not affect the power of the generated sneaker flaticon, but only its overall temporal duration.

In the case of four frequency jumps, the collision of the resulting four individual subcritical flaticons always leads to their merging and formation of an individual, high power sneaker flaticon. The critical case of panel (a) in Fig.\ref{coll4} is once again obtained for a total jump amplitude $f_j=2.4f_c$. Indeed, the resulting steady pulse has a larger temporal width than the corresponding critical pulse of Fig.\ref{coll2} because of the longer initial temporal extension of the ensemble of colliding flaticons. Note that the peak powers of the merged flaticons may once again be well predicted by means of Eq.(\ref{rop}): for example, in the supercritical case of panel (c) in Fig.\ref{coll4}, Eq.(\ref{rop}) predicts the peak power of 3.4 W, which is in good agreement with the numerical value of about 3.2 W.  

\begin{figure}[ht]
\centering
\includegraphics[width=4cm]{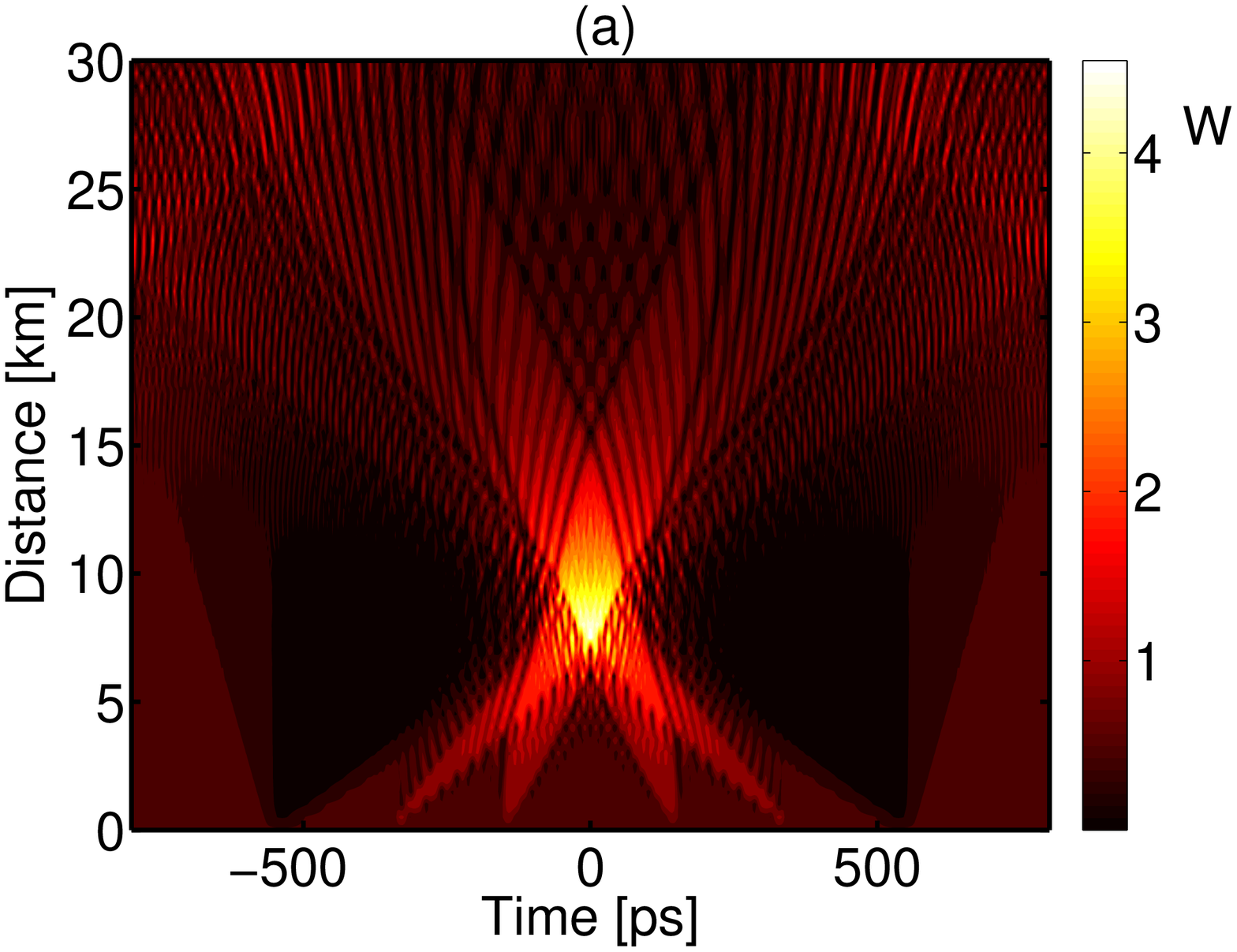}
\includegraphics[width=4cm]{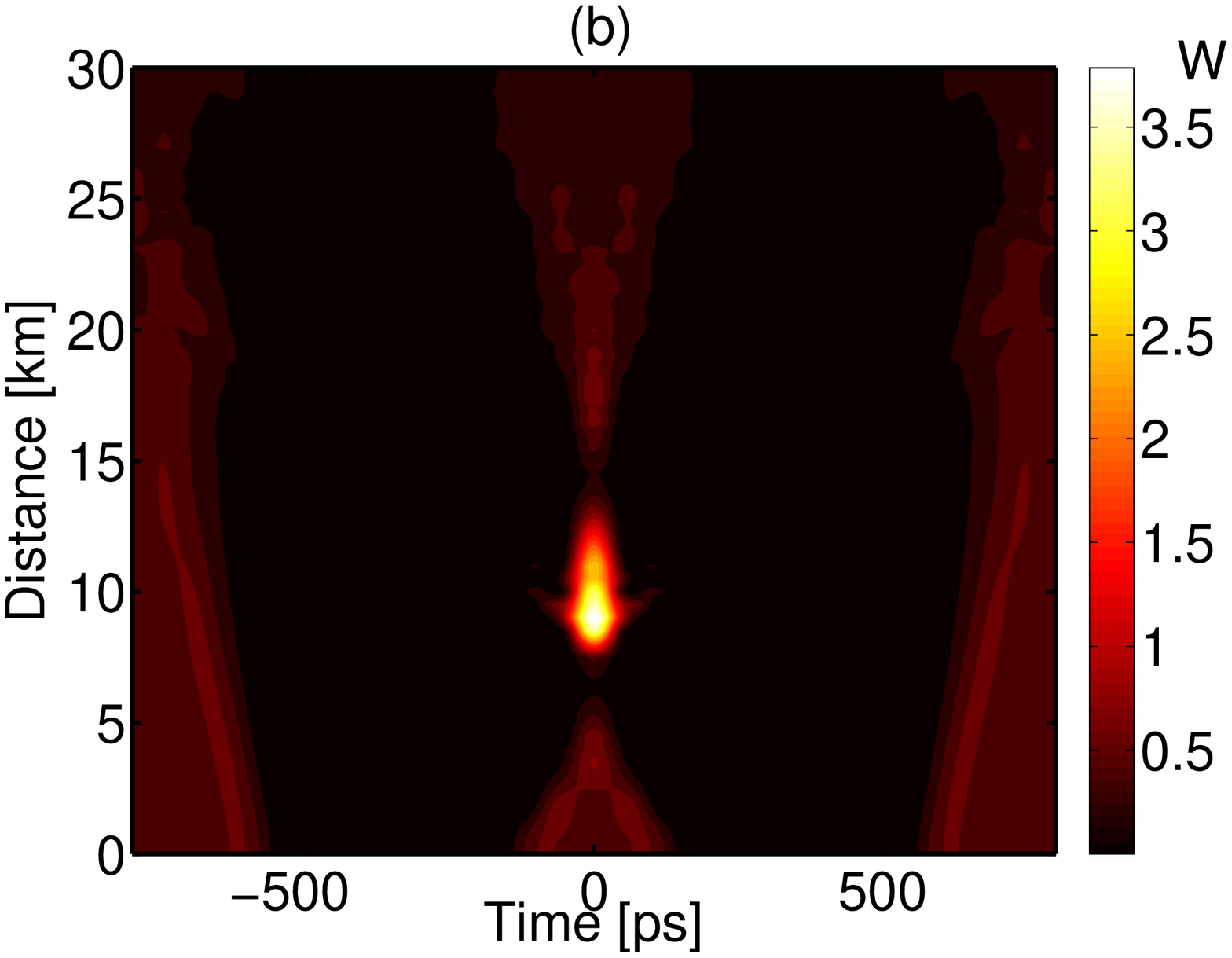}  
\includegraphics[width=5cm]{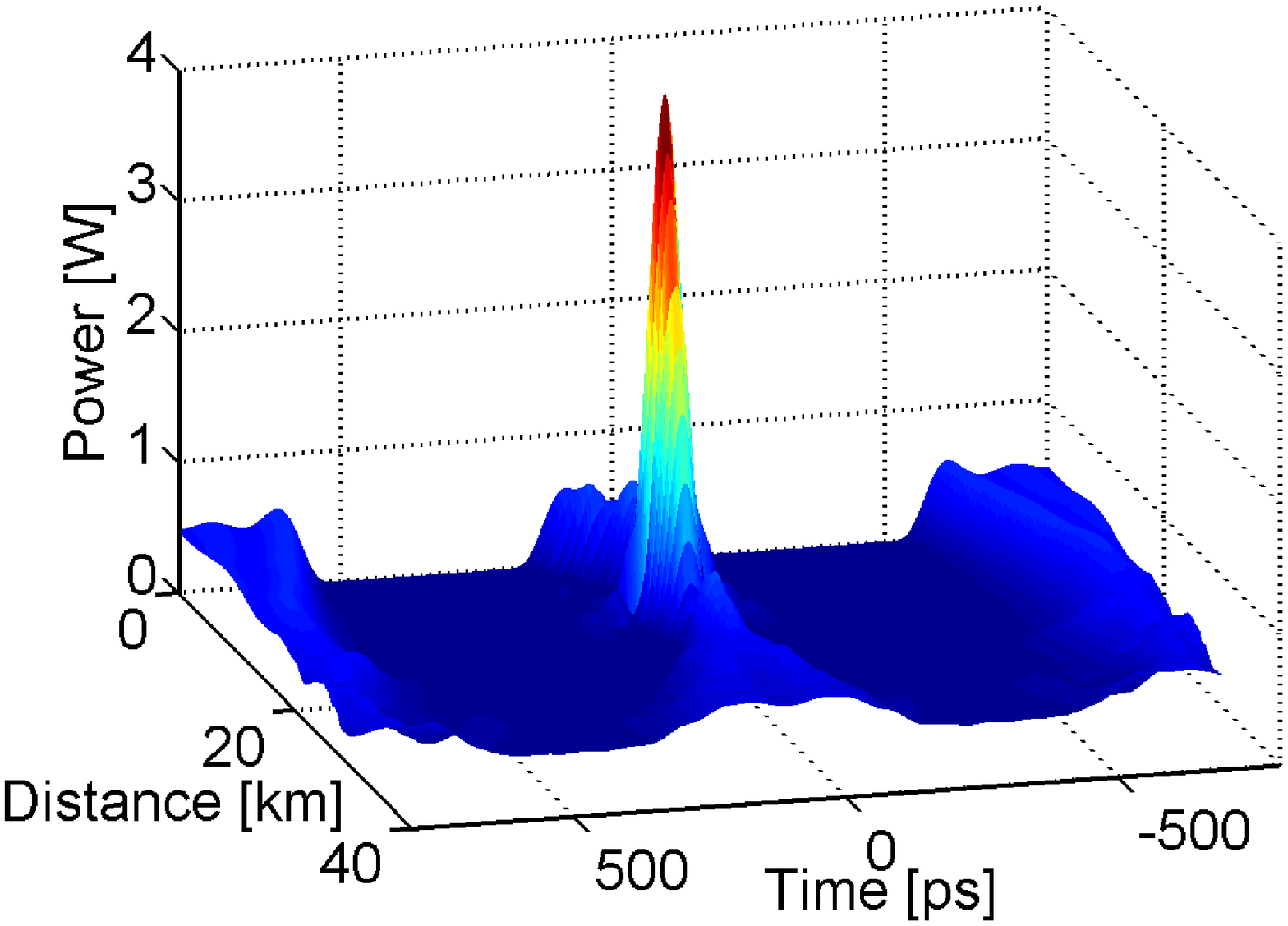}  
\caption{(a) Contout plot of the optical power from the collision of four flaticons with a supercritical total frequency jump $f_j=4f_c$; (b) same as in the left panel, after bandpass filtering; (c) surface plot corresponding to panel (b)}
\label{coll4bis}
\end{figure}

Even larger peak powers may be obtained by further increasing the amplitude of the individual subcritical frequency jumps, so that a collective supercritical jump results. For example, in Fig.\ref{coll4bis} we illustrate the case of a total jump of amplitude $f_j=4f_c$. In this case Eq.(\ref{rop}) predicts that the peak power of the collision-generated flaticon is equal to 4.5 W, that is nine times higher than the input CW laser power level, in excellent agreement with the numerical simulation shown in panel (a) of Fig.\ref{coll4bis}. 

The price to be paid for obtaining peak powers larger than the critical value $\rho_c$ is that both the temporal duration and spatial extension of the collision-induced flaticon is progressively reduced as the peak power grows larger. The finite lifetime or flash-like nature of the supercritical flaticons further justifies their classification among the class of deterministic rogue waves. In fact, by increasing the number N of individual subcritical colliding flaticons that are generated by frequency jumps with half amplitude, say, $u_s$, it is possible obtain a collective flaticon of arbitrarily high peak power, equal to \cite{biondini06}

\begin{equation}\label{fpower}
	\rho_{max}=\rho_0\left(1+\frac{N u_s}{2\sqrt{\rho_0}}\right)^2
\end{equation}

The spatio-temporal localization of the high amplitude pulse resulting from the collision of the four flaticons is more clearly visualized in panels (b) and (c) of Fig.\ref{coll4bis}, where we display the contour plot and the surface plot of the evolution with distance of the optical power, after passing the total field through the same bandpass filter as in Fig.\ref{sjump}. Note that the bandpass filter is merely used here to more clearly visualize the flash-like nature of the collision-induced supercritical flaticon by removing the high-frequency wave-breaking oscillations, but it does not affect its localization in both space and time, nor its peak power value, as can be seen by comparing panels (a) and (b) of Fig.\ref{coll4bis}.  

The flaticon collision-induced generation of a collectively supercritical rogue wave may also be observed in optical communication systems using the wavelength-division-multiplexing (WDM) technique, associated with an intensity modulation format (such as NRZ). Consider for example a dispersion-managed, long distance, periodically amplified transmission link with a net or average normal GVD: in this case the NLSE (\ref{nls2}) represents the path-averaged propagation. In such situation, the dynamics which is observed in Figs.\ref{coll4}-\ref{coll4bis} corresponds to collisions among sequences of equal power and initially in-phase and non-overlapping pulses in five different channels, with the relative frequency spacing of $2u_s$. Namely, the input condition shown in Figs.\ref{coll4} would correspond to the sequence $(0100000)$ in the channel at relative frequency $+4u_s$, the sequence $(0010000)$ at frequency $+2u_s$, the sequence $(1001001)$ at the central frequency, the sequence $(0000100)$ at frequency $-2u_s$, and the sequence $(0000010)$ at $-4u_s$, respectively. The formation of a rogue peak with power up to $(1+N)^2$ times the power of the individual channel pulses in a WDM system comprising $N$ channels (where $N$ may be as large as $N=100$) may lead to damages in peak power-sensitive devices, and result in significant error bursts, for example whenever the peak is produced at the receiver position. In actual WDM communication systems, the statistics of such collision-induced rogue peaks will directly reflect the statistical distribution of pulse sequences in the various channels \cite{sergei}. In order to avoid such extreme wave events, it should be convenient to use phase-modulated as opposed to intensity modulated formats.     

\section{Periodic rogue wavetrains}
\label{sec:wave}

\begin{figure}[ht]
\centering
\includegraphics[width=6cm]{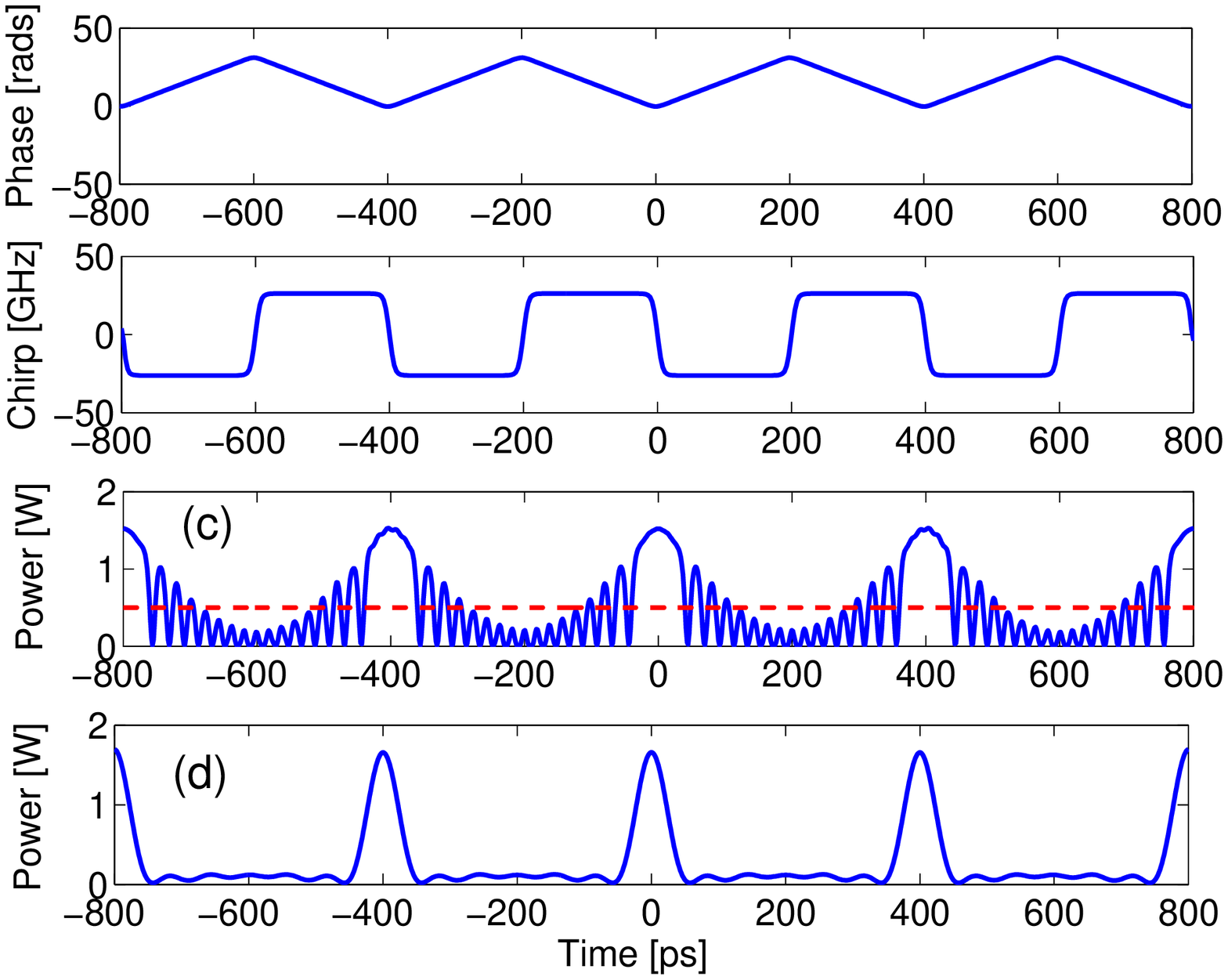}
\includegraphics[width=4cm]{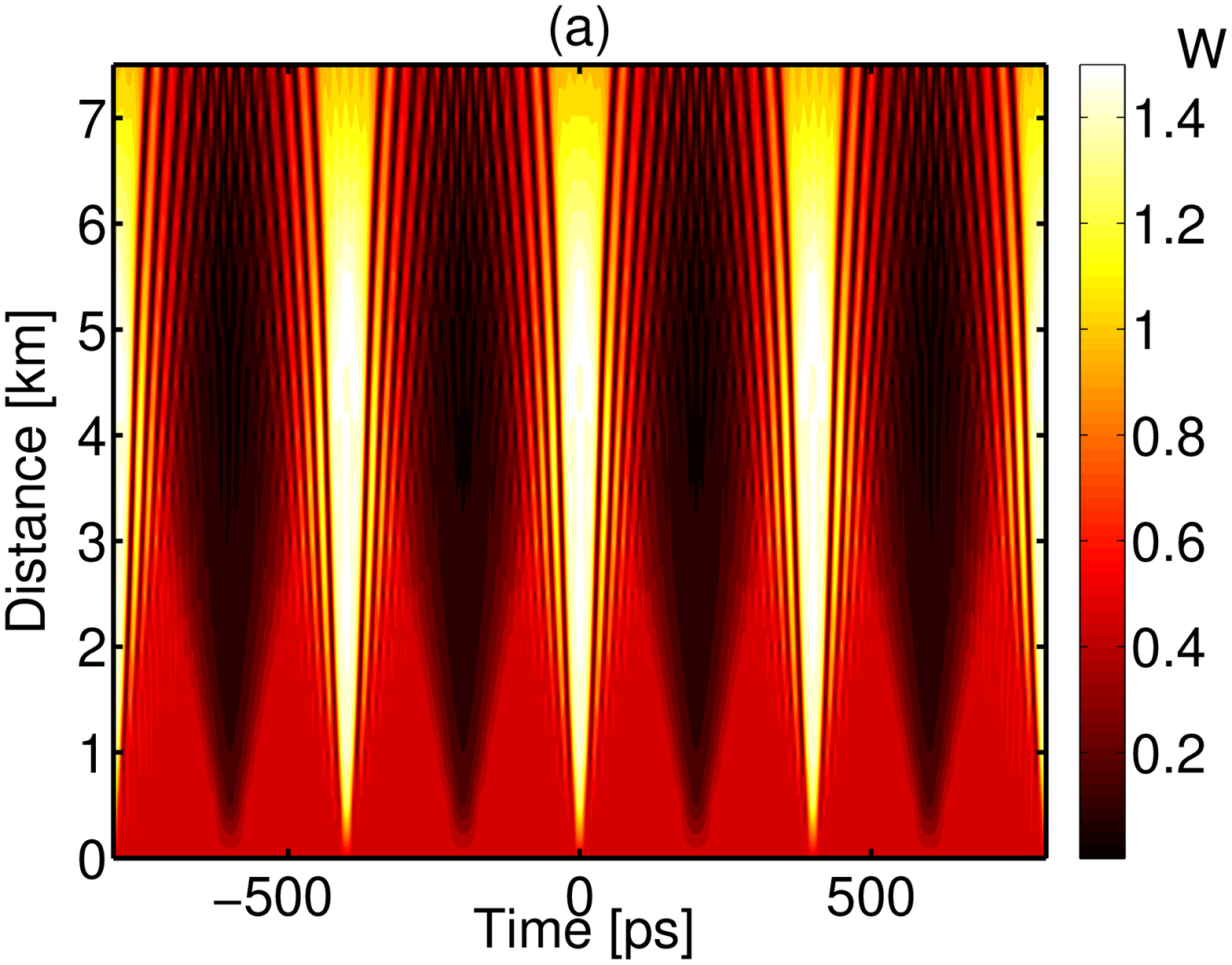} 
\includegraphics[width=4cm]{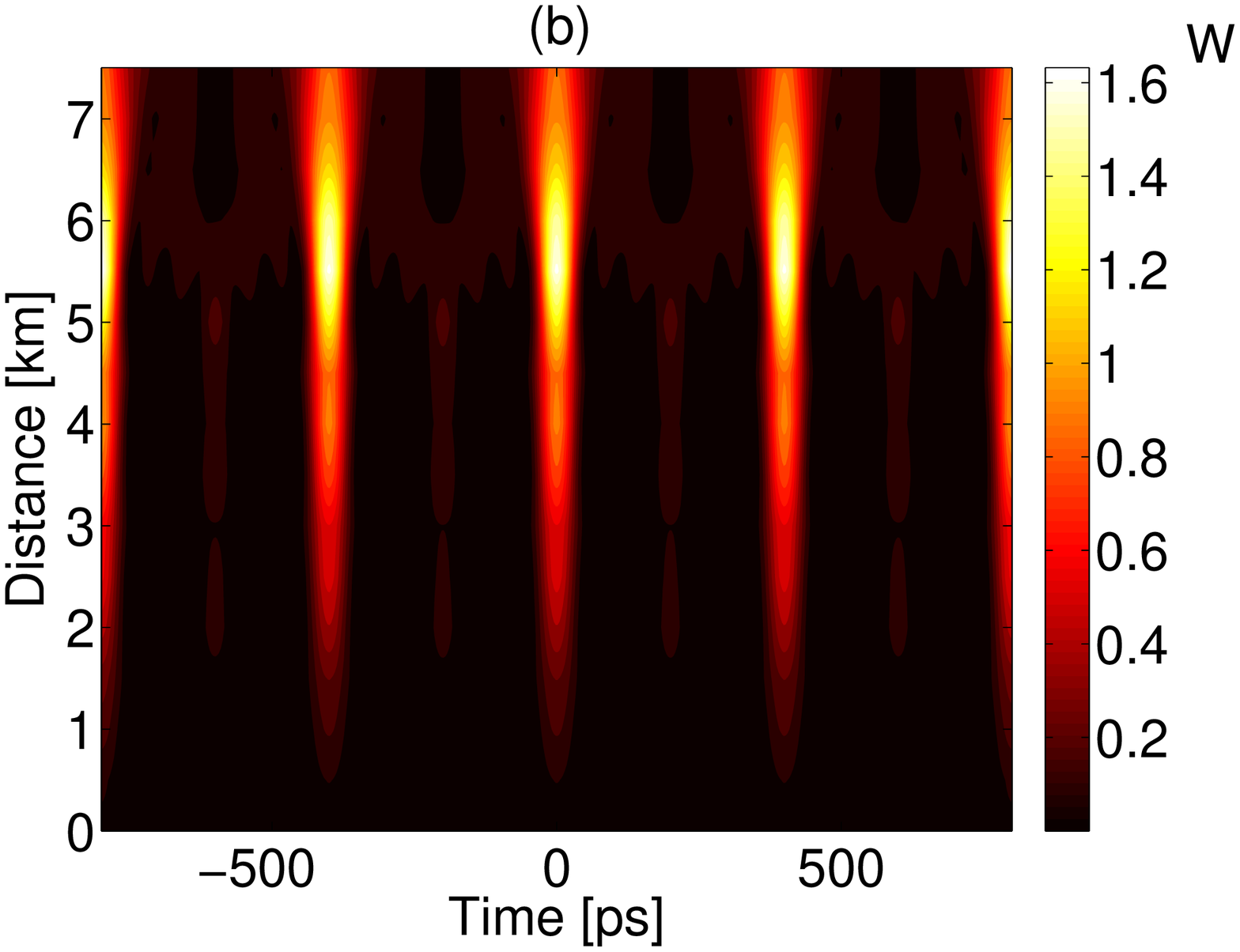}
\caption{Top: Input temporal phase and chirp evolution of 2.5 GHz frequency modulated CW laser and (c) unfiltered or (d) filtered output pulse train profile at 5.5 km of DCF; Bottom: Power vs. DCF length ($D=-100\: ps/nm\cdot km$) for $P=500 \:mW$: a) total power; (b) power after bandpass filtering.}  
\label{periodic}
\end{figure}

In previous sections we have studied the dynamics of individual flaticons, induced by a single or multiple input frequency jumps. Let us consider now the possible application of optical flaticons to the generation of periodic high intensity pulse trains in nonlinear optical fibers. For example, in Fig.\ref{periodic} we show the case of an input frequency modulation with a period of 400 ps (i.e., 2.5 GHz). The resulting evolution along the fiber of the optical power of the full field is shown in panel (a) of Fig.\ref{periodic}. As it can be seen, a periodic train of flaticon pulses is generated, whose temporal width $T_p$ grows larger with distance according to the law of Eq.(\ref{tp}). The corresponding power profile after bandpass filtering (we used here again the same filter as in Fig.\ref{sjump}) is shown in panel (b) of Fig.\ref{periodic}: the maximum peak power of about 1.7 W for the generated 2.5 GHz pulse train is obtained after 5.5 km of DCF (see panels (c) and (d), before and after filtering, respectively) and is characterized by a full width at half maximum after filtering of 51 ps (which corresponds to the duty cycle of $12.75\%$) . 

From the practical point of view, it would be useful to replace the initial triangular phase modulation of the CW laser by a simple sinusoidal modulation. Indeed, our study of flaticon collisions in section (\ref{sec:colli}) shows that, whenever the number N of frequency jumps grows larger, and correspondingly their individual amplitude $f_s$ is reduced (so that the total frequency jump remains equal to the critical value $f_j=Nf_s=2.4f_c$), the corresponding phase and frequency modulation (see Fig.\ref{coll4}) may approach in the continuous limit a sinusoidal waveform.

\begin{figure}[ht]
\centering
\includegraphics[width=6cm]{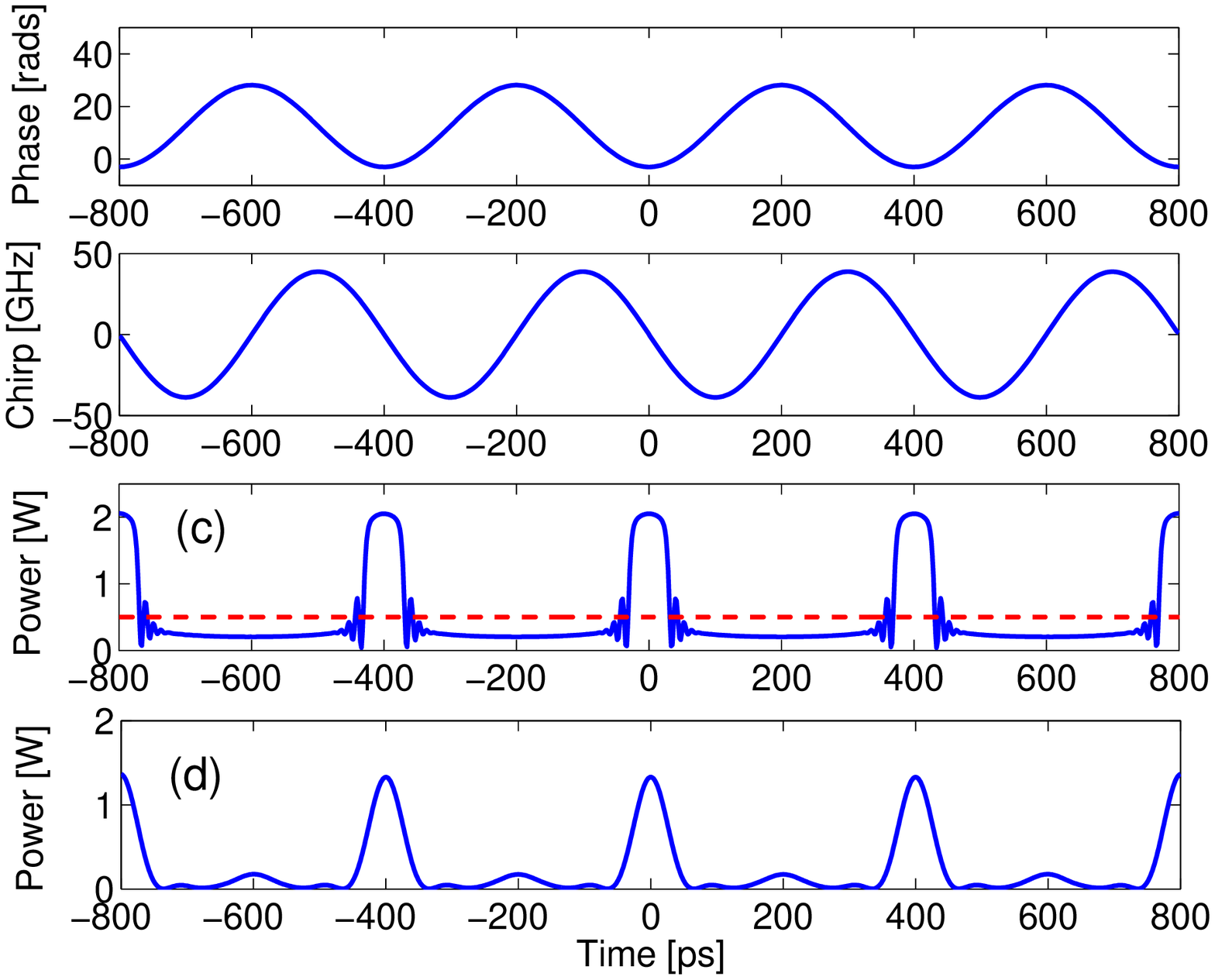}
\includegraphics[width=4cm]{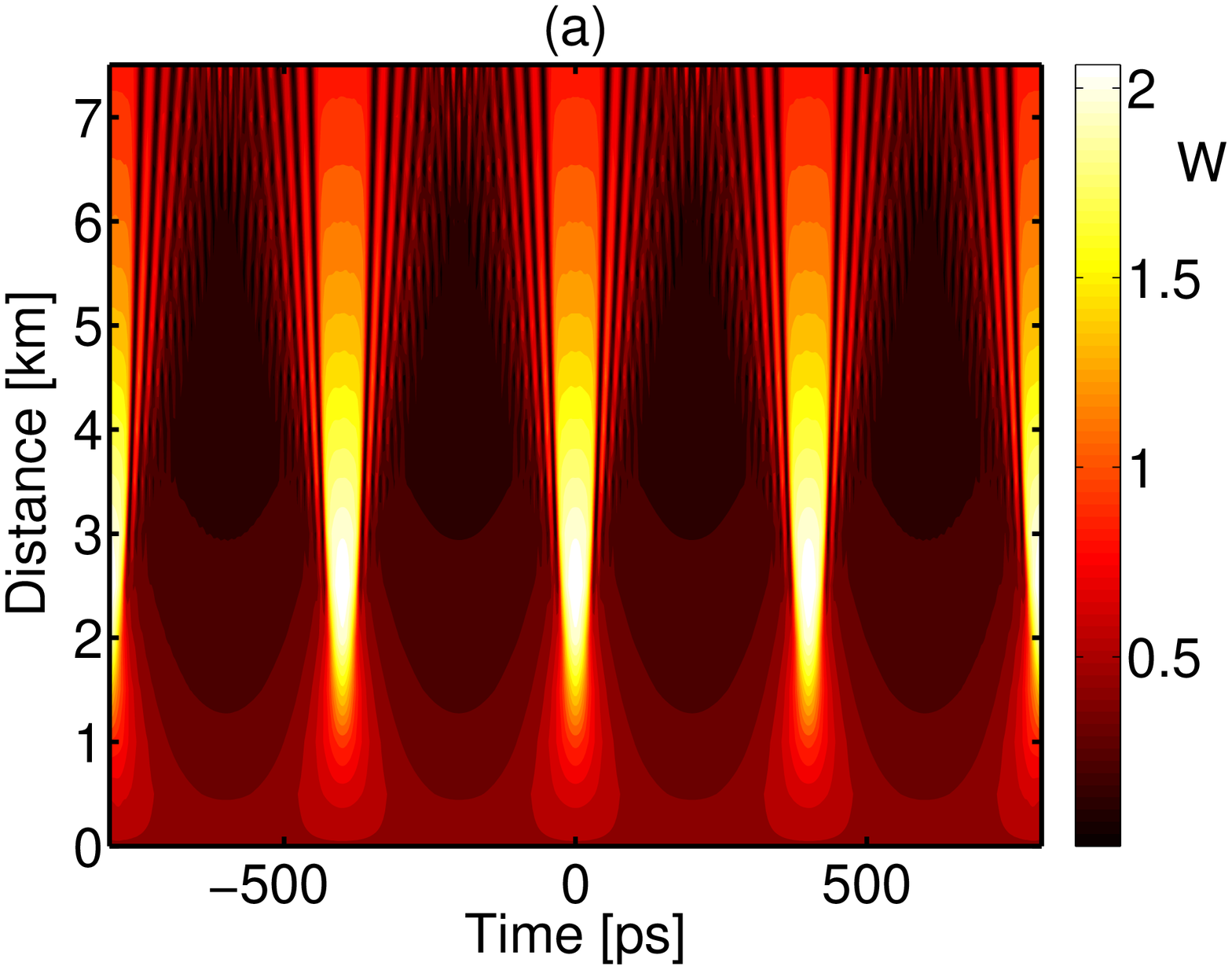}  
\includegraphics[width=4cm]{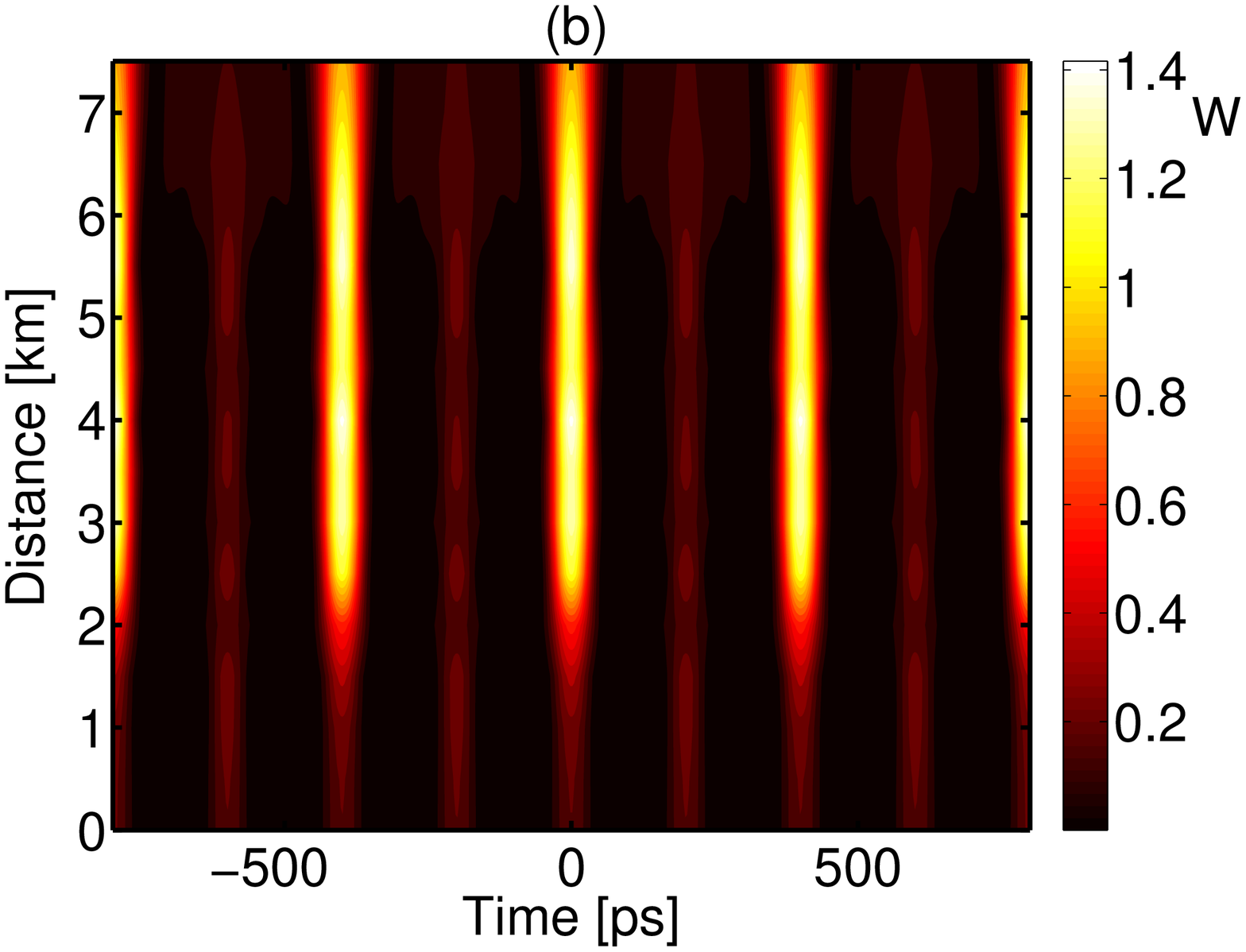} 
\caption{As in Fig. \ref{periodic}, with input sinusoidal phase modulation. (a) total power; (b) power after bandpass filtering; (c) unfiltered or (d) filtered output pulse train profile at 3 km of DCF; }
\label{periodicsin}
\end{figure}

Therefore we have numerically studied the generation of a pulse train where the triangular phase modulation of Fig.\ref{periodic} is replaced by a pure sinusoidal phase modulation, where the peak value of the phase modulation is kept the same in both cases. The contour plot of the (a) panel in Fig. \ref{periodicsin} indeed shows that a flaticon pulse train is generated with sinusoidal phase modulation as well, albeit with a gradually decreasing peak power after the first 3 km of DCF. 

On the other hand, the contour plot of the (b) panel of Fig.\ref{periodicsin} which is obtained after bandpass filtering (using the same bandpass filter as in Fig.\ref{periodic}) shows quite strikingly that the generation of a low duty ratio train of pulses is even enhanced whenever the periodic sinusoidal phase modulation replaces triangular phase modulation. Indeed, the high-intensity and high-extinction ratio train of pulses is obtained after the first 3 km of DCF already with sinusoidal modulation, as opposed to 5.5 km with triangular phase modulation (see panels (c) (before filtering) and (d) (after filtering) of Fig.\ref{periodicsin}): here the full width at half maximum is equal to 56 ps after filtering. Moreover, panel (b) of Fig.\ref{periodicsin} also shows that with sinusoidal phase (and frequency) modulation, the power of the generated pulse train remains relatively stable with distance, as opposed to the case of Fig.\ref{periodic} where the optimal fiber length should be carefully adjusted.

Although in our analysis we neglected the presence of linear fiber loss in order to better highlight the essentials of the pulse dynamics, we verified that the dynamics of flaticon and rogue wave generation remains qualitatively unaffected when fiber loss is included. In practice, in short fiber spans it is possible to fully compensate linear loss by means of distributed Raman amplification. Whereas in long distance transmission links loss is periodically compensated by means of optical amplifiers: in such a case, the NLSE (\ref{nls2}) has the meaning of a guiding center or path-averaged propagation equation, which also applies to dispersion-managed systems \cite{yuji95}.

\section{Conclusions}
\label{sec:concl}
We have described the conditions for the practical observation of extreme wave events in the MI-free regime of normal GVD of optical fibers. Such rogue waves are obtained through the collision of self-similar waveforms or flaticons resulting from the initial phase or frequency modulation of a CW laser, or from the nonlinear superposition of pulse patterns in dispersion-managed wavelength division transmission links. In analogy with shallow water sneaker waves induced by the merging of currents moving in opposite directions along the coastline, optical shallow water rogue waves appear as flashes of high-intensity, flat-top and chirp-free humps of light. These waves have, on the one hand, beneficial potential application to time-compressed and high intensity optical pulse train generation. On the other hand, they may lead to information loss and catastrophic damages in intensity-modulated, dispersion managed WDM transmission links. In any case, optical fibers may provide the ideal testbed for exploring the dynamics of these extreme waves which may occur in diverse fields of application such as oceanography, geophysics, ferromagnetics, plasmas, etc. An interesting perspective for further studies is the extension of these collision-induced rogue waves to the spatial domain, that is replacing dispersion by diffraction, where the two-dimensional degrees of freedom may be exploited for simplifying the generation and control of the initial spatial phase profile \cite{wan}. Finally, analogous phenomena may appear in Bose-Einstein condensation where the NLSE reduces to the NSWE in the semiclassical limit, i.e., whenever the Planck's constant $\hbar$ is a small parameter with respect to the quantities to be measured.

\section{Acknowledgments}
This work was carried out with support from the Conseil Regional de Bourgogne, and the iXCore Foundation. The present research is supported in Brescia by Fondazione Cariplo, grant n.2011-0395.











%
%

\end{document}